\documentclass[aps,nofootinbib,tightenlines]{revtex4}
\usepackage{epsfig}
\usepackage{amsfonts}
\usepackage{amsmath}
\usepackage{graphicx}
\usepackage{color}
\usepackage{amssymb,amsmath,psfrag,slashed,graphicx}
\newcommand{\beq}{\begin{equation}}
\newcommand{\eeq}{\end{equation}}
\newcommand{\bqa}{\begin{eqnarray}}
\newcommand{\eqa}{\end{eqnarray}}

\def \non {\nonumber}

\def \t {\tilde}

\begin{document}

\title{Gluon quasidistribution function at one loop }

\author{Wei Wang$^1$~\footnote{wei.wang@sjtu.edu.cn}, Shuai Zhao$^1$~\footnote{shuai.zhao@sjtu.edu.cn}, Ruilin Zhu$^{2,1}$~\footnote{rlzhu@njnu.edu.cn} }
\affiliation{
$^1$ INPAC, Shanghai Key Laboratory for Particle Physics and Cosmology, School of Physics and Astronomy, Shanghai Jiao Tong University, Shanghai, 200240,   China\\
$^2$ Department of Physics and Institute of Theoretical Physics,
Nanjing Normal University, Nanjing, Jiangsu 210023, China }

\begin{abstract}
We study  the unpolarized gluon quasidistribution function in the
nucleon at one loop level in the large momentum effective theory. For the quark quasidistribution,  power law ultraviolet divergences arise  in the cut-off scheme and  an important observation is that  they all are subjected  to   Wilson lines. However for  the gluon quasidistribution function,  we first point out  that the linear  ultraviolet
divergences also exist in  the real diagram  which is  not connected to  any Wilson
line. We then study  the one loop corrections to
parton distribution functions  in both cut-off scheme and  dimensional regularization   to deal with the ultraviolet
divergences.  In addition to the ordinary quark and gluon distributions, we also include  the quark to gluon and gluon to quark splitting  diagrams. The complete one-loop matching factors between the  quasi and light cone parton distribution functions  are  presented in the cut-off scheme.  We derive  the $P^z$ evolution equation for  quasi parton
distribution functions, and find  that the $P^z$ evolution kernels are identical to the DGLAP evolution kernels.

\end{abstract}

\maketitle

\section{Introduction}

Hadron structures are determined by strong interactions among quarks and gluons. In order to make predictions for high energy process involving hadrons, one must rely the factorization approach.
Parton distribution functions (PDFs) are important inputs to
evaluate the hadron scattering cross section in the collinear factorization.   However, up to date it is still  a formidable  task to calculate the PDF from first principles because QCD are only well understood
in the perturbative regime of short distance phenomena, whereas the long distance interactions require  a nonperturbative approach.

The only  method that  allows   to  properly explore the
nonperturbative interactions  is Lattice QCD (LQCD). PDFs are defined as  matrix elements of non-local
operators located on the light cone. Thus they are time-dependent, and  can not be directly calculated  on LQCD. With the help of  operator product expansion,   PDFs can be expressed in terms of  a series of moments that are calculable on LQCD in principle. However only the lowest few moments can be computed on LQCD, while the calculation for higher moments is much more involved due to the operator mixing and the noisy signal.

A more realistic and phenomenological  approach is  often adopted at this stage.
One may  appeal to some theoretical parametrization of   PDFs and  fit the free parameters in the parametrization  using the experiment data such as DIS and $pp$ collisions data~\cite{Butterworth:2015oua,Hou:2016nqm,Hou:2017khm}. Such an approach has reached very high accuracy, for instance one has taken into account the next-to-next-to-leading order in $\alpha_s$ for the calculation of partonic cross section.  This
method keeps the dependence parametrization of PDFs, and also can not cover all the region of
the parton longitudinal momentum fraction $x$.

In 2013, a new approach named as the large momentum effective theory
(LaMET)
was proposed~\cite{Ji:2013dva,Ji:2014gla} to determine parton distribution functions on the lattice. In this approach, instead of using
the correlations of quarks and gluons on the light cone, one makes use of   the matrix elements of frame-dependent equal-time
correlators in the large momentum limit. The   equal-time
correlators, defining the quasi parton distribution functions,  can  be
directly investigated  on the Lattice QCD. In the $P^z\to \infty$ limit,  this effective theory allows us to perform a matching between the ordinary PDFs and the quasi ones. Through this way,   the  light-cone PDFs can be elegantly extracted. Moreover in fact this approach can generally  allow one to derive   the information of the
light cone parton physics from first principles of QCD.

This approach has been developed rapidly, and a number of progresses
have been made
recently~\cite{Xiong:2013bka,Ma:2014jla,Ma:2014jga,Ji:2014lra,Ji:2015qla,Xiong:2015nua,Monahan:2015lha,Jia:2015pxx,Li:2016amo,Li:2017udt,Nam:2017gzm,
Radyushkin:2016hsy,Radyushkin:2017gjd,Zhang:2017bzy,Radyushkin:2017ffo,Carlson:2017gpk,Briceno:2017cpo,Xiong:2017jtn,
Radyushkin:2017cyf,Constantinou:2017sej,Rossi:2017muf,Orginos:2017kos,Ji:2017rah,Broniowski:2017wbr}.
A key issue in this approach concerns the renormalizability  of the
quasi
PDF~\cite{Ji:2015jwa,Chen:2016fxx,Monahan:2016bvm,Alexandrou:2017huk,Chen:2017mzz,Ji:2017oey,Ishikawa:2017faj,Green:2017xeu}.
Using the heavy quark and Wilson line correspondence,  an all order
formal proof has been presented in Ref.~\cite{Ji:2017oey}. More
explicitly, the one loop matching between nonsinglet quark
distributions and quasidistributions has been studied in
Refs.~\cite{Xiong:2013bka,Ma:2014jga}, in which  the linear power
law ultraviolet (UV) divergences have been found in the cut-off
scheme. To  absorb the linear divergences, an approach using  the
nondipolar Wilson line has been proposed in
Ref.~\cite{Li:2016amo,Li:2017udt}; the other one is Wilson line
renormalization by including a ``mass''
counterterm~\cite{Chen:2016fxx,Ishikawa:2016znu,Ji:2017oey,Ishikawa:2017faj}.
On the other side,  the direct evaluation of the parton distribution
functions have also been performed on the lattice in
Refs.~\cite{Lin:2014zya,Alexandrou:2015rja,Chen:2016utp,Bacchetta:2016zjm,Alexandrou:2017huk,Chen:2017mzz,Bali:2017ude,Yoon:2017qzo},
and the accuracy  is incredibly increasing.

It has been recognized for a  long time that the gluon distribution functions are one of the key
input parameters for the physics at hadron colliders. In  the production of
Higgs boson and heavy quarkonium, contributions from gluon PDFs are dominant over the quark ones~\cite{Butterworth:2015oua,Hou:2016nqm,Hou:2017khm}.  For these reasons, it is mandatory to explore the quasi PDF for the gluon in the LaMET, which is still not available yet in the literature. The purpose of this paper is to fill this gap.  To do so,   we will calculate the perturbative function of gluon
quasidistribution at one loop level. Both the cut-off  and
dimensional regularization scheme are employed in our calculation.
We will discuss the difficulties in the application of UV cut-off
scheme, and all of the matching functions will be presented in
dimensional regularization (DR). The    splitting
functions are also studied. Based on the matching functions, we derive  the
$P^z$ evolution equation for the parton quasidistributions.

The rest of this paper is organized as follows. In Sec.~\ref{II}, we discuss the one
loop results for gluon quasidistribution and light cone
distribution  in the cut-off scheme. In Sec.~\ref{sec:1loop}, we give the one
loop matching results for parton quasidistributions and light cone
distributions  at the dimensional regularization scheme (DR).  The $P^z$ evolution equation of
parton quasidistributions is also presented.   We summarize   in the  last section. In the appendix,  some calculation details in the cut-off scheme and dimensional regularization are  relegated to Sec.~\ref{sec:appendix_cutoff} and Sec.~\ref{sec:appendix_dr}, respectively. In both schemes, we have used the finite gluon mass to regulate the infrared divergences, while  in the appendix \ref{sec:appendix_offshell}, we present  the results with the offshellness as an infrared regulator.

\section{Gluon quasidistribution function in LaMET\label{II}}
\subsection{Parton distribution functions}

PDFs   are important inputs to
evaluate the hadron scattering cross section in high energy physics. In the case of unpolarized quark distribution in a hadron,
the light cone distribution is defined as
\begin{eqnarray}\label{qlcDef}
   f_{q/H}(x,\mu^2) &=& \int \frac{d\xi^-}{4\pi} e^{-i\xi^- xP^+}  \langle P|\overline{\psi}(\xi^-)
   \gamma^+   {\cal P} \exp\left(-ig\int^{\xi^-}_0 d\eta A^+(\eta) \right)\psi(0) |P\rangle,
\end{eqnarray}
where $H$ denotes the hadron, and $|P\rangle$ is the hadron state with
momentum $P$, $x= k^+/P^+$ is the longitudinal momentum fraction.

When discussing light-cone distributions we adopt light-cone
coordinate, while for quasi PDFs, the ordinary Minkowski
coordinates are used. In the light-cone coordinate system,  a vector
$a$ is expressed as
$a=(a^+,a^-,\vec{a}_{\perp})=((a^0+a^3)/\sqrt{2},
(a^0-a^3)/\sqrt{2}, a^1, a^2)$. We also have $k\cdot p= k_\mu p^\mu=
g_{\mu\nu}k^\mu p^\nu= p^+ k^- + p^- k^+ -\vec{p}_\perp \cdot
\vec{k}_\perp$. We introduce two unit light-cone vectors
$n_+^\mu=(0,1,0,0)$, $n_-^{\mu}=(1,0,0,0)$ in the light-cone coordinate system, and the transverse metric
$g^{\mu\nu}_{\perp}$. In light-cone coordinate system,
$g^{\mu\nu}_{\perp}=g^{\mu\nu}-n_+^{\mu}n_-^{\nu}-n_-^{\mu}n_+^{\nu}$,
and in ordinary Minkowski coordinates it can be written as
$g^{\mu\nu}_{\perp}=g^{\mu\nu}+n_z^{\mu}n_z^{\nu}-n_0^{\mu}n_0^{\nu}$,
where $n_z=(0,0,0,-1)$ and $n_0=(1,0,0,0)$. The nucleon momentum in
the light-cone coordinates is written as
$P^\mu=(P^+,M^2/(2P^+),0,0)$. The above operator in Eq.~\eqref{qlcDef} is
non-local and time-dependent.

In the case of unpolarized
gluon distribution in the nucleon, the corresponding light cone distribution is  defined as~\cite{Collins:1981uw}
\begin{eqnarray}\label{gUnpolDef}
   x  f_{g/H}(x,\mu^2) &=& \int \frac{d\xi^-}{2\pi P^+} e^{-i\xi^- xP^+}  \langle P|G^+_{~\mu}(\xi^-)
   {\cal P}\exp\left(-ig\int^{\xi^-}_0 d\eta A^+(\eta) \right)G^{\mu +}(0) |P\rangle \ .
\end{eqnarray}
 The covariant derivation is defined as $D_\mu =\partial_\mu + i g A_\mu$. Then $G_{\mu\nu}=T^a G^a_{\mu\nu}=-\frac{i}{g}[D_\mu,D_\nu]=T^a (\partial_\mu A^a_\nu-\partial_\nu A^a_\mu-g f^{abc}A^b_\mu A^c_\nu)$.

In LaMET,  the quasidistributions  are introduced, which can
be calculated directly on the lattice.  The unpolarized quark quasidistribution is defined as~\cite{Ji:2013dva}
\begin{eqnarray}\label{qUnpolDef}
   \tilde f_{q/H}(x,\mu^2, P^z) &=& \int \frac{dz}{4\pi} e^{izxP^z}  \langle P|\overline{\psi}(z)
   \gamma^z  {\cal P} \exp\left(-ig\int^{z}_0 dz'_z A^z(z') \right)\psi(0) |P\rangle \ ,
\end{eqnarray}
where $x= k^z/P^z$ with the nucleon momentum $P^z$. All the fields
are at $\xi^0=t=0$ and lie along the $\xi^3=z$ direction.
Correspondingly, the $\gamma^+$ is replaced by $\gamma^z$ in the
expression. The above quantity is non-local and time-independent,
which can be simulated on the lattice for any $P^z\ll 1/a$, where
$a$ is the lattice spacing.

According to the similar extension, the unpolarized
gluon quasidistribution can be defined as~\cite{Ji:2013dva,Ma:2014jga}
\begin{eqnarray}\label{gUnpolDef}
   x \tilde  f_{g/H}(x,\mu^2, P^z) &=& \int \frac{dz}{2\pi P^z} e^{izxP^z}  \langle P|G^z_{~\mu}(z)
   {\cal P}\exp\left(-ig\int^{z}_0 d\eta A^z(\eta) \right)G^{\mu z}(0) |P\rangle,
\end{eqnarray}
where $\mu$ sums over the transverse components.
The   parton gauge link (Wilson line) is defined along the $z$ direction
\begin{eqnarray}
W(z_2,z_1)=\mathcal{P}\exp\left(-ig\int^{z_2}_{z_1} dz' A^z(z')\right).
\end{eqnarray}
 For gluon
quasidistribution, the gauge link is in adjoint representation.
Through the definitions of PDFs, one can obtain
\begin{align}
  f_{q/H}(x)=- f_{\bar q/H} (-x),~~\t f_{q/H}(x)=-\tilde f_{\bar q/H}(-x),
\end{align}
and
\begin{align}
f_{g/H}(x)=-f_{g/H}(-x),~~ \t f_{g/H}(x)=-\t
f_{g/H}(-x),\label{eq:supportprops}
\end{align}
by exchanging the two quark or gluon fields in the operator
definitions. $\bar q$ and $\t{\bar q}$ are the light-cone and quasi
distributions of anti-quark.

In the infinite momentum  limit, the nucleon can be viewed as a beam
of free partons. Employing the language of quantum field theory, the
hard parts of parton quasidistributions and standard distributions
can be calculated order by order. The nonperturbative parts of
quasidistributions can be calculated on the lattice. Thus we can
evaluate the light cone distributions through the matching condition
between the quasidistributions and standard distributions.  In LaMET, the
standard distribution and quasidistribution are connected by the
factorization formula
\begin{eqnarray}
\t f_{i/H}(x, P^z)=\int^{1}_{0}\frac{dy}{y}
  Z_{ij}\left(\frac{x}{y},\frac{\mu}{P^z}\right)f_{j/H}(y, \mu).
\end{eqnarray}

\subsection{Feynman rules for gluon quasidistribution function}

\begin{figure}[hbt]
\begin{center}
\includegraphics[width=0.5\textwidth]{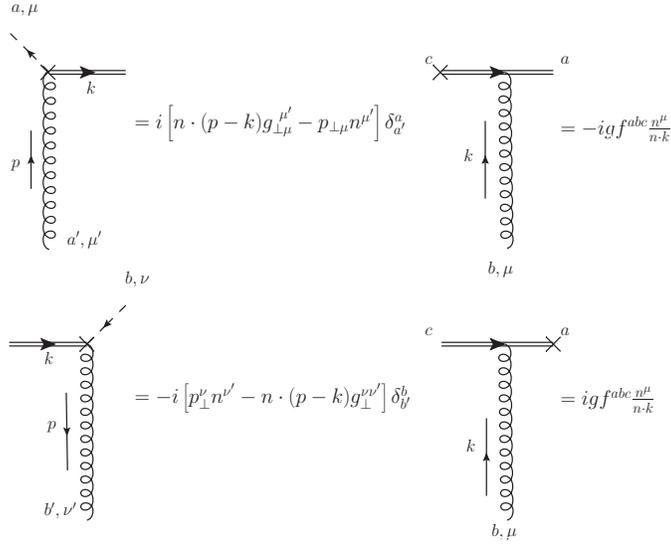}
\end{center}
\caption{Feynman rules for quasi gluon distribution. }
\label{fig:feynrules}
\end{figure}

In the following, we will calculate the gluon quasidistributions  in Feynman gauge.
The gauge link $W(z,0)$ can be divided into two parts
\begin{eqnarray}
W(z,0)=W^{\dag}(+\infty,z)W(+\infty,0),
\end{eqnarray}
then a half of the matrix element can be written as
\begin{eqnarray}\label{w1}
W_{da}(+\infty,z_1)F_{\mu,a}^{~z}(z_1)&=&W_{da}(+\infty,z_1)(\partial_{\mu}A^z_a-\partial^zA_{a,\mu}
-gf_{abc}A_{b,\mu}A^{z}_c).
\end{eqnarray}
With the help of integration by parts, the second and the third
term of the right side can be rewritten in a compact form
\begin{eqnarray}
-\partial^zA_{a,\mu}
-gf_{abc}A_{b,\mu}A^{z}_c=-\partial^z(W_{ad}(+\infty,z_1)A^d_{\mu}(z_1)),
\end{eqnarray}
then Eq.~\eqref{w1} becomes
\begin{eqnarray}\label{w2}
W_{da}(+\infty,z_1)F_{\mu,a}^{~z}(z_1)&=&W_{da}(+\infty,z_1)\partial_{\mu}A^z_a -\partial^z(W_{da}(+\infty,z_1)A_{a,\mu}).
\end{eqnarray}
It is obvious that the coupling between the abelian part of field
strength tensor and gauge link, and the non-Abelian part of field strength tensor
can be reorganized and described by one single Feynman rule.
Expanding  Eq.\eqref{w2} to $\mathcal{O}(g^1)$ in the momentum space
will directly lead to the Feynman rule, which is shown in the left
panel of Fig.~\ref{fig:feynrules}. The coupling between eikonal line
and gluon is shown in the right panel of Fig.~\ref{fig:feynrules}.

\subsection{Cut-off scheme}

\begin{figure}[htbp]
\centering
\includegraphics[width=0.6\textwidth]{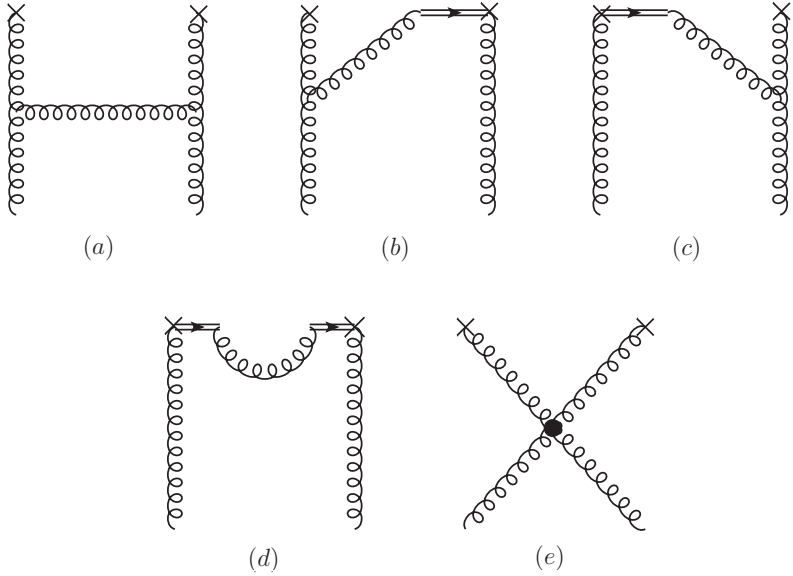}
\caption{One-loop real corrections to quasi (light-cone) gluon
distribution in Feynman gauge. The double line includes both the
gauge link and non-abelian contributions.} \label{fig:gluonreal}
\end{figure}

\begin{figure}[htbp]
\centering
\includegraphics[width=0.7\textwidth]{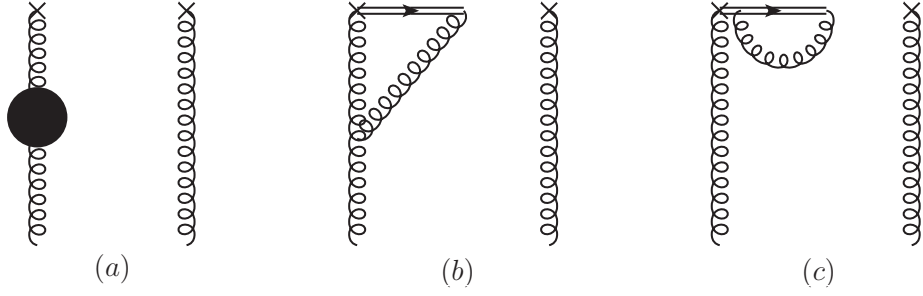}
\caption{One-loop virtual corrections to quasi (light-cone) gluon
distribution in Feynman gauge, the conjugate diagrams are not shown.
The double line includes both the gauge link and non-abelian
contributions.} \label{fig:gluonvirtual}
\end{figure}

 In Feynman gauge, the one-loop diagrams for gluon distribution are depicted in
Fig.~\ref{fig:gluonreal} and Fig.~\ref{fig:gluonvirtual}. The
vertexes from gauge link and non-abelian contributions are
incorporated. The gluon mass is
introduced to regulate the infrared divergences, and we will use
the cut-off scheme to regulate the UV divergences.

The one-loop correction can be generally expressed as
\begin{eqnarray}
\tilde f_{g/g}(x, \Lambda, P^z) &=& (1+\delta\tilde
{\mathcal{C}}^{(1)}_{gg}(\Lambda, P^z)+\dots) \delta(x-1) + \tilde
f_{g/g}^{(1)}(x,\Lambda, P^z) + \dots ,
\end{eqnarray}
where $\delta\tilde{\mathcal{C}}^{(1)}_{gg}(\Lambda, P^z)$ accounts
for the virtual correction and $\tilde f_{g/g}^{(1)}(x,\Lambda,
P^z)$ denotes the real correction.

The contribution from Fig.~\ref{fig:gluonreal}(a) reads
\begin{widetext}
\begin{align}
x\t f_{g/g}^{(1)}\bigg\vert_{\mathrm{Fig}. \ref{fig:gluonreal}(a)}=&\int\frac{d^4k}{(2\pi)^4}(-gf_{a_1 b_1 c_1}[(P+k)^{\gamma_1} g^{\mu_1 \nu_1}+(-2k+P)^{\mu_1}g^{\nu_1 \gamma_1}+(k-2P)^{\nu_1}g^{\mu_1 \gamma_1}])\non\\
&
\times (-gf_{a_2 b_2 c_1}[(-P-k)^{\gamma_2} g^{\mu_2 \nu_2}+(2k-P)^{\mu_2}g^{\nu_2 \gamma_2}+(2P-k)^{\nu_2}g^{\mu_2 \gamma_2}])\non\\
&\times (i)(k\cdot n g^{\nu'_2}_{\perp\mu}
-n^{\nu'_2}k_{\perp\mu})(-i)(n^{\nu'_1}k^{\mu}_{\perp} -k\cdot n
g^{\nu'_1\mu}_{\perp})\delta^{b_1}_{b_2}
\frac{\delta(x-k^z/P^z)}{ 2(N^2_c-1)(P^z)^2}\non\\
&\times \frac{-ig_{\nu_2 \nu'_2}}{k^2-m_g^2}\frac{-ig_{\nu_1
\nu'_1}}{k^2-m_g^2}\frac{-ig_{\gamma_1
\gamma_2}}{(P-k)^2-m_g^2}(-g_{\perp \mu_1
\mu_2}\delta^{a_1}_{a_2}). \label{eq:gluonreal:amp1}
\end{align}
\end{widetext}
A straightforward calculation yields the quasi PDF
\begin{align}
 \tilde f_{g/g}^{(1)}(x, P^z, \Lambda)\bigg\vert_{\mathrm{Fig}.\ref{fig:gluonreal}(a)}&=\frac{\alpha_s C_A}{2\pi x}\left\{ \begin{array}
{ll}\left(2 x^3-3 x^2+2 x-2\right) \ln \frac{x-1}{x}+2 x^2-\frac{5
x}{2}+\frac83+\frac34\frac{\Lambda}{P^z}\ , & x>1
\\ \left(2 x^3-3
x^2+2 x-2\right)\ln \frac{(x^2-x+1)m_g^2}{4x(1-x)(P^z)^2}+\frac{1}{6} x \left(4 x (8 x-9)-\frac{9}{(x-1) x+1}+42\right)\\
-\frac{8}{3}+\frac34\frac{\Lambda}{P^z}\ , &0<x<1\\
-\left(2 x^3-3 x^2+2 x-2\right) \ln \frac{x-1}{x}-2 x^2+\frac{5
x}{2}-\frac83+\frac34\frac{\Lambda}{P^z}\ ,&x<0\end{array}
\right.\label{eq:quasi:cutoff}
\end{align}
and the light-cone PDF
\begin{align}
f_{g/g}^{(1)}(x,
\Lambda)\bigg\vert_{\mathrm{Fig}.\ref{fig:gluonreal}(a)}&=\frac{\alpha_s
C_A}{2\pi x}\left\{ \begin{array} {ll}  0\ , & x>1\  \mbox{or}\ x<0
\\ \left(2 x^3-3 x^2+2
x-2\right) \ln \frac{\left(x^2-x+1\right)m_g^2}{\Lambda^2}-\frac{3
x}{2(x^2-x+1)}+2x^2(x-1)+\frac{7}{2}x-2\ ,&0<x<1 \end{array}
\right.\label{eq:lc:cutoff}
\end{align}
In the above,  $\tilde f_{g/g}$ and $f_{g/g}$ are quasi and
light-cone distributions respectively. One can find  that, the
infrared divergences, which are regularized as logarithms of
$m_g^2$, are the same in Eqs.~\eqref{eq:quasi:cutoff} and
\eqref{eq:lc:cutoff}. It indicates   that the quasi and light-cone
distributions have the same infrared structure, and the LaMET
factorization holds for the gluon case at one-loop level. The quasi
distribution has non-zero values in the region $x>1$ and $x<0$,
while the light-cone distribution is zero in these regions. The
infrared divergence of quasi distribution only exists in the region
$0<x<1$.

A few remarks on the above results are given in  order.
\begin{itemize}

\item From the amplitude for Fig.~\ref{fig:gluonreal}(a) given in Eq.~\eqref{eq:gluonreal:amp1},  one can see that the gluon field strength tensor contributes with a factor $k$, and the three gluon vertex also contributes with a $k$. The $k^z$ integration is determined by the $\delta$ function, and thus this amplitude is proportional to
\begin{eqnarray*}
\int  d^3k \frac{k^4}{k^6} \sim \frac{\Lambda}{P^z}.
\end{eqnarray*}
For the light-cone PDF, one finds that this linear divergence is proportional to $n_+^2$ which vanishes.

\item
One can notice that, in light-cone distribution, the UV
divergence is logarithmic, while quasi distribution suffers a linear
power UV divergence. This will bring an obstacle
in the Lattice calculation and need to be properly renormalized. In
fact, the linear power divergences have already been discussed  in quasi quark
distribution, but they  only show up  in the real diagram where the two pieces of
gauge link are connected by a gluon line. Based on this fact, a small
mass counter term of gauge link is introduced to deal with the
divergence \cite{Chen:2016fxx,Ji:2017oey,Ishikawa:2016znu}. However,
in the gluon case, the  linear power divergence appears in
Eq.~\eqref{eq:quasi:cutoff} and  has nothing to do with the Wilson line.  It indicates that linear power UV
divergence in gluon case must  be treated in a different approach.
For Figs.~\ref{fig:gluonreal}(b) and (c) where the gluon gauge
link participate in, the linear divergence exists as well.
Results for these diagrams will be arranged to appendix A.

\item
A more severe  difficulty in applying  the UV cut-off  scheme to quasi gluon
distribution comes from   self-energy diagram
Fig.~(\ref{fig:gluonvirtual}a). It is well known that the
cut-off scheme for the vacuum polarization will suffer quadratic
divergence, even in QED. The source of this divergence is that an UV
cut-off will break the gauge symmetry.   In the
following, we will perform our calculation in
DR scheme. However, it would be valuable  to find out a cut-off scheme respecting the gauge
invariance when calculating in lattice perturbation theory.

\item We have checked that if the index  $\mu$ also runs over the time component, i.e. $\mu=0, 1,2$, the results for the quasi-PDF  are linearly divergent as well.

\item We should note that there are also contributions from the
``crossed'' diagrams, which can be obtained by exchanging the two gluon
lines entering the non-local vertex. They give non-zero
contributions to light-cone distribution in the region $-1<x<0$, and
for quasi distribution, they contribute to the region $x>0$,
$-1<x<0$, and $x<-1$.  The ``crossed'' diagrams and their
contributions are not explicitly shown  in this work, since their contributions  can be easily derived  by using
Eq.~\eqref{eq:supportprops}.
\end{itemize}

\section{One-loop Results for parton quasidistributions and light cone distributions in Dimensional
Regularization}\label{sec:1loop}

In the literature, the matching of quark quasi-PDF has been performed in both cut-off and dimensional regularization schemes. The calculation in $\overline{\mathrm{MS}}$ scheme is meaningful to the discussions on the multiplicative renormalizability of quasi-PDF \cite{Ji:2015jwa} and the nonperturbative renormalization in RI/MOM scheme \cite{Alexandrou:2017huk,Chen:2017mzz,Green:2017xeu,Lin:2017ani,Stewart:2017tvs}.
For these reasons we think that it is also valuable to perform perturbative calculation in DR scheme for gluon quasi PDF, just like the case for quark quasi distribution. Besides, it is also convenient to derive the $P^z$ evolution equations in DR scheme.

We also note that lattice collaborations had already performed a few calculations on the lattice and some promising results are obtained \cite{Lin:2014zya,Alexandrou:2015rja}, without a renormalization of the linear divergence. Therefore, even the problem of linear ultraviolet  divergences is unsolved at present, studying the matching of gluon quasidistribution function in $\overline{\mathrm{MS}}$ scheme shall still provide some information to simulate the gluon PDF on the lattice.

In the following we will present our one-loop calculation for quasi and light-cone distributions of quark and gluon. In the calculation,  we will work in the Feynman gauge and adopt the dimensional
regularization   to regularize the UV divergence. The  infrared divergence will be regularized by a small
gluon mass $m_g$.

\subsection{Quark in quark}

\begin{figure}[htbp]
\includegraphics[width=0.8\textwidth]{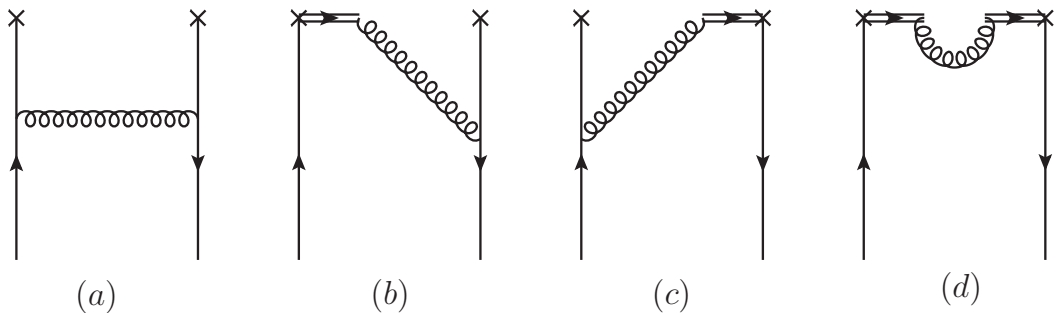}
\caption{One-loop real corrections to  quark quasi (light-cone)
distributions in Feynman gauge. } \label{fig:quarkreal}
\end{figure}
\begin{figure}[htbp]
\includegraphics[width=0.6\textwidth]{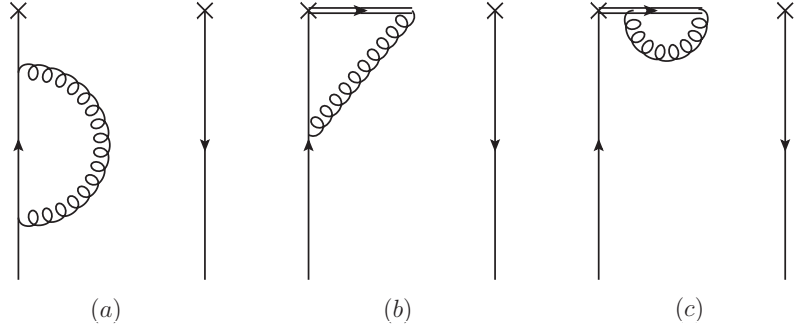}
\caption{One-loop virtual corrections to  quark quasi (light-cone)
distributions in Feynman gauge, where the conjugate diagrams are not
shown. }\label{fig:quarkvirtual}
\end{figure}
The Feynman diagrams for one-loop corrections to quark in quark
distributions  are shown in Figs.~\ref{fig:quarkreal} and
\ref{fig:quarkvirtual}, where the diagrams in
Fig.~\ref{fig:quarkreal} are the real corrections and the ones in
Fig.~\ref{fig:quarkvirtual} are the virtual corrections.
Similar with  gluon distributions, there are also ``crossed''
diagrams, which can be obtained  by exchanging the two quark lines
connecting to the non-local vertex. These diagrams can be explained as
one-loop corrections to quasi and light-cone distributions of
anti-quark. According to Eq.~\eqref{eq:supportprops}, one can
immediately recover the anti-quark distributions.

We will  calculate the one-loop real corrections to quasi and
light-cone quark distributions diagram by diagram.
Therein, Fig.~\ref{fig:quarkreal}(a) gives  the amplitude
\begin{align}
\tilde f_{q/q}^{(1)}(x,\mu)\bigg\vert_{\mathrm{Fig}.
\ref{fig:quarkreal}(a)}&=\mu^{2\epsilon}\int\frac{d^{4-2\epsilon}
k}{(2\pi)^{4-2\epsilon}}\bar u(P)(-ig
T^a\gamma^\mu)\frac{i}{\slashed k}\gamma^z\frac{i}{\slashed
k}(-igT^a\gamma_\mu)\frac{-i}{(P-k)^2-m_g^2}u(P)\frac{\delta(x-\frac{k^z}{P^z})}{
4N_cP^z},
\end{align}
which leads to
\begin{eqnarray}
\tilde f_{q/q}^{(1)}(x,
P^z)\bigg\vert_{\mathrm{Fig}.\ref{fig:quarkreal}(a)}=\frac{\alpha_s
C_F}{2\pi}\left\{\begin{aligned}
&(x-1)\ln\frac{x-1}{x}+1,~~&x>1\\
&(x-1)\ln\frac{m_g^2}{4(1-x)(P^z)^2}+x,~~&0<x<1\\
&(x-1)\ln\frac{x}{x-1}-1,~~&x<0
\end{aligned}
\right.
\end{eqnarray}
For the light-cone PDF, we have
\begin{eqnarray}
f_{q/q}^{(1)}(x,\mu)\bigg\vert_{\mathrm{Fig}.\ref{fig:quarkreal}(a)}=\frac{\alpha_s
C_F}{2\pi}\left\{
\begin{aligned}
&0,~~&x>1,~~x<0\\
&(x-1)\left(\ln\frac{m_g^2 x}{{\mu}^2}+2\right),~~&0<x<1
\end{aligned}
\right.
\end{eqnarray}

The second and third diagrams have the identical result
\begin{align}\label{gamma2}
\tilde f_{q/q}^{(1)}(x,\mu)\bigg\vert_{\mathrm{Fig}.
\ref{fig:quarkreal}(b,c)}&=\mu^{2\epsilon}\int\frac{d^{4-2\epsilon}
k}{(2\pi)^{4-2\epsilon}}\bar u(P)
(igT^a)\frac{-i}{n\cdot(P-k)}\gamma^z\frac{i}{\slashed k}(-ig
T^a\gamma^z)\frac{-i}{(P-k)^2-m_g^2}u(P)\frac{\delta(x-\frac{k^z}{P^z})}{
4N_cP^z}.
\end{align}
The contributions to quasi PDFs are derived as:
\begin{eqnarray}
&&\tilde f_{q/q}^{(1)}(x,
P^z)\bigg\vert_{\mathrm{Fig}.\ref{fig:quarkreal}(b)}=\tilde
f_{q/q}^{(1)}(x,
P^z)\bigg\vert_{\mathrm{Fig}.\ref{fig:quarkreal}(c)} =\frac{\alpha_s
C_F}{4\pi}\left\{
\begin{aligned}
&\frac{2x}{1-x}\ln\frac{x}{x-1}-\frac{1}{1-x},~~~~&x>1\\
&\frac{2x}{1-x}\ln\frac{4(1-x)(P^z)^2}{m_g^2}+\frac{1-2x}{1-x},~~~~&0<x<1\\
&\frac{2x}{1-x}\ln\frac{x-1}{x}+\frac{1}{1-x},~~~~&x<0
\end{aligned}\right.
\end{eqnarray}
While for the light-cone PDF, we have
\begin{eqnarray}
&&f_{q/q}^{(1)}(x,\mu^2)\bigg\vert_{\mathrm{Fig}.\ref{fig:quarkreal}(b)}=f_{
q/q}^{(1)}(x,\mu^2)\bigg\vert_{\mathrm{Fig}.\ref{fig:quarkreal}(c)}
=\frac{\alpha_s C_F}{2\pi}\left\{
\begin{aligned}
&0,~~~~&x>1,~x<0\\
&\frac{x}{1-x}\ln\frac{\mu^2}{m_g^2 x},~~~~&0<x<1
\end{aligned}
\right.
\end{eqnarray}

Fig.~\ref{fig:quarkreal}(d) gives
\begin{align}\label{gamma13}
\tilde f_{q/q}^{(1)}\bigg\vert_{\mathrm{Fig}.
\ref{fig:quarkreal}(d)}&=\mu^{2\epsilon}\int\frac{d^{4-2\epsilon}
k}{(2\pi)^{4-2\epsilon}}\bar u(P)\gamma^z (ig T^a)(-ig T^a)n^2
\frac{-i}{n\cdot(P-k)}\frac{i}{n\cdot(P-k)}\frac{-i}{(P-k)^2-m_g^2}
u(P)\frac{\delta(x-\frac{k^z}{P^z})}{ 4N_cP^z},
\end{align}
which corresponds to
\begin{eqnarray}
\t
f_{q/q}^{(1)}(x)\bigg\vert_{\mathrm{Fig}.\ref{fig:quarkreal}(d)}=\frac{\alpha_s
C_F}{2\pi}\left\{
\begin{aligned}
&\frac{1}{1-x},~~~~&x>1\\
&\frac{1}{x-1},~~~~&0<x<1\\
&\frac{1}{x-1},~~~~&x<0
\end{aligned}
\right.
\end{eqnarray}
The light-cone PDF receives no contribution from this diagram since $n^2\equiv n^2_+=0$:
\begin{eqnarray}
f_{q/q}^{(1)}(x,\mu^2)\bigg\vert_{\mathrm{Fig}.\ref{fig:quarkreal}(d)}=0.
\end{eqnarray}

Next we will calculate the virtual corrections, where their conjugate diagrams contributions are included. The quark self energy diagram will give
\begin{eqnarray}
&&\delta\widetilde{\mathcal{C}}_{qq}^{(1)}\bigg\vert_{\mathrm{Fig}.\ref{fig:quarkvirtual}(a)}
=-\frac{\alpha_s C_F}{2\pi}\int dy\left\{\begin{aligned}
&(y-1)\ln\frac{y-1}{y}+1,~~&y>1\\
&(y-1)\ln\frac{m_g^2}{4(1-y)(P^z)^2}+y,~~&0<y<1\\
&(1-y)\ln\frac{y-1}{y}-1,~~~~&y<0
\end{aligned}\right.
\end{eqnarray}
and
\begin{eqnarray}
\delta
\mathcal{C}^{(1)}_{qq}\bigg\vert_{\mathrm{Fig}.\ref{fig:quarkvirtual}(a)}=-\frac{\alpha_s
C_F}{2\pi}\int^1_0 dy~y\left( \ln\frac{\mu^2}{m_g^2 y}-1\right).
\end{eqnarray}

Fig.~\ref{fig:quarkvirtual}(b) gives
\begin{align}\label{gamma2}
\t {\mathcal{C}}^{(1)}_{qq}\bigg\vert_{\mathrm{Fig}.
\ref{fig:quarkvirtual}(b)}&=\mu^{2\epsilon}\int\frac{d^{4-2\epsilon}
k}{(2\pi)^{4-2\epsilon}}\bar u(P)
(-igT^a)\frac{-i}{n\cdot(P-k)}\gamma^z\frac{i}{\slashed k}(-ig
T^a\gamma^z)\frac{-i}{(P-k)^2-m_g^2}u(P)\frac{\delta(x-1)}{
4N_cP^z},
\end{align}
and thus
\begin{eqnarray}
&&\delta\widetilde{\mathcal{C}}^{(1)}_{qq}\bigg\vert_{\mathrm{Fig}.\ref{fig:quarkvirtual}(b)}
=-\frac{\alpha_s C_F}{2\pi}\int dy\left\{
\begin{aligned}
&\frac{2y}{1-y}\ln\frac{y}{y-1}-\frac{1}{1-y},~~&y>1\\
&\frac{2y}{1-y}\ln\frac{4(1-y)(P^z)^2}{m_g^2}+\frac{1-2y}{1-y},~~&0<y<1\\
&\frac{2y}{1-y}\ln\frac{y-1}{y}+\frac{1}{1-y},~~&y<0
\end{aligned}\right.
\end{eqnarray}
The  corrections for light-cone PDF are
\begin{eqnarray}
\delta
\mathcal{C}^{(1)}_{qq}\bigg\vert_{\mathrm{Fig}.\ref{fig:quarkvirtual}(b)}=-\frac{\alpha_s
C_F}{2\pi}\int^1_0 dy \frac{2y}{1-y}\ln\frac{\mu^2}{m_g^2 y}.
\end{eqnarray}

Fig.~\ref{fig:quarkvirtual}(c) gives the amplitude
\begin{align}\label{gamma2}
\t {\mathcal{C}}^{(1)}_{qq}\bigg\vert_{\mathrm{Fig}.
\ref{fig:quarkvirtual}(c)}&=\mu^{2\epsilon}\int\frac{d^{4-2\epsilon}
k}{(2\pi)^{4-2\epsilon}}\bar u(P)\gamma^z (-ig T^a)(-ig T^a)n^2
\frac{-i}{n\cdot(P-k)}\frac{i}{n\cdot(P-k)}\frac{-i}{(P-k)^2-m_g^2}
u(P)\frac{\delta(x-1)}{ 4N_cP^z}.
\end{align}
Corrections to quasi-PDF are then derived as
\begin{eqnarray}
\delta\widetilde{\mathcal{C}}^{(1)}_{qq}\bigg\vert_{\mathrm{Fig}.\ref{fig:quarkvirtual}(c)}=-\frac{\alpha_s
C_F}{2\pi}\left\{
\begin{aligned}
&\frac{1}{1-y},~~~~&y>1\\
&\frac{1}{y-1},~~~~&0<y<1\\
&\frac{1}{y-1},~~~~&y<0
\end{aligned}
\right.
\end{eqnarray}
and for the light-cone, it is again vanishing
\begin{eqnarray}
\delta
\mathcal{C}^{(1)}_{qq}\bigg\vert_{\mathrm{Fig}.\ref{fig:quarkvirtual}(c)}=0.
\end{eqnarray}

\subsection{Gluon in gluon}

The Feynman diagrams for one-loop corrections to gluon quasi and
light-cone distributions have been shown in Fig.~\ref{fig:gluonreal} and
Fig.~\ref{fig:gluonvirtual}, where the diagrams in
Fig.~\ref{fig:gluonreal} are for real corrections and the diagrams
in Fig.~\ref{fig:gluonvirtual} are for virtual corrections.

For the real corrections, we have
\begin{align}
 \tilde f_{g/g}^{(1)}(x, P^z)\bigg\vert_{\mathrm{Fig}.\ref{fig:gluonreal}(a)}&=\frac{\alpha_s C_A}{2\pi x}\left\{ \begin{array}
{ll}\left(2 x^3-3 x^2+2 x-2\right) \ln \frac{x-1}{x}+2 x^2-\frac{5
x}{2}+\frac83\ , & x>1
\\ \left(2 x^3-3
x^2+2 x-2\right)\ln \frac{(x^2-x+1)m_g^2}{4x(1-x)(P^z)^2}+\frac{x}{6} \left(4 x (8 x-9)-\frac{9}{(x-1) x+1}+42\right)-\frac{8}{3}\ , &0<x<1 \\
-\left(2 x^3-3 x^2+2 x-2\right) \ln \frac{x-1}{x}-2 x^2+\frac{5
x}{2}-\frac83\ ,&x<0 \end{array} \right.\label{eq:gluonreal1:result}
\end{align}
and
\begin{align}
f_{g/g}^{(1)}(x,
\mu)\bigg\vert_{\mathrm{Fig}.\ref{fig:gluonreal}(a)}&=\frac{\alpha_s
C_A}{2\pi x}\left\{ \begin{array} {ll}  0\ , & x>1\  \mbox{or}\ x<0\
\\ \left(2 x^3-3 x^2+2
x-2\right) \ln \frac{\left(x^2-x+1\right)m_g^2}{\mu^2}-\frac{3
x}{2(x^2-x+1)}+2x^2(x-1)+\frac{7}{2}x-2\ ,&0<x<1  \end{array}
\right.
\end{align}

Fig.~\ref{fig:gluonreal}(b)  yields
\begin{align}
x\t f_{g/g}^{(1)}\bigg\vert_{\mathrm{Fig}.
\ref{fig:gluonreal}(b)}=&\mu^{2\epsilon}\int\frac{d^{4-2\epsilon}
k}{(2\pi)^{4-2\epsilon}}(-gf_{a_1 b_1 c_1}[(P+k)^{\gamma_1} g^{\mu_1
\nu_1}+(-2k+P)^{\mu_1}g^{\nu_1 \gamma_1}+(k-2P)^{\nu_1}g^{\mu_1
\gamma_1}])\non\\&\times(i)(k\cdot n g^{~\mu_2}_{\perp\mu}
-n^{\mu_2}P_{\perp\mu})(-i)(n^{\nu'_1}k^{\mu}_{\perp} -k\cdot n
g^{\nu'_1 \mu}_{\perp})
\frac{\delta(x-k^z/P^z)}{ 2(N^2_c-1)(P^z)^2}\frac{-i}{(P-k)\cdot n }(-gf_{a_2 c_1 b_1} n^{\gamma_2})\non\\
&\times\frac{-ig_{\nu_1 \nu'_1}}{k^2-m_g^2}\frac{-ig_{\gamma_1
\gamma_2}}{(P-k)^2-m_g^2}(-g_{\perp\mu_1 \mu_2}\delta^{a_1}_{a_2}),
\end{align}
which leads to
\begin{eqnarray}
 \tilde f_{g/g}^{(1)}(x, P^z)\bigg\vert_{\mathrm{Fig}.\ref{fig:gluonreal}(b,c)} =-\frac{\alpha_s C_A}{4\pi}\left\{
\begin{aligned}
&\frac{1}{x-1}\left[x(1+x)\ln\frac{x}{x-1}-2x+1\right],~~~~&x>1\\
&\frac{1}{x-1}\left[x(1+x)\ln\frac{4x(1-x)(P^z)^2}{(1-x+x^2)m_g^2}-(2x^2-2x+1)\right],~~~~&0<x<1\\
&\frac{1}{x-1}\left[x(1+x)\ln\frac{x-1}{x}+2x-1\right],~~~~&x<0
\end{aligned}\right.
\end{eqnarray}
and
\begin{eqnarray}
f_{g/g}^{(1)}(x,
\mu)\bigg\vert_{\mathrm{Fig}.\ref{fig:gluonreal}(b)}=
f_{g/g}^{(1)}(x,
\mu)\bigg\vert_{\mathrm{Fig}.\ref{fig:gluonreal}(c)}=-\frac{\alpha_s
C_A}{4\pi}\left\{
\begin{aligned}
&0,~~~~&x>1,~x<0\\
&{\frac{x(1+x)}{x-1}\ln\frac{\mu^2}{m_g^2(1-x+x^2)}},~~~~&0<x<1
\end{aligned}
\right.
\end{eqnarray}

Fig.~\ref{fig:gluonreal}(d) gives
\begin{align}
x\t f_{g/g}^{(1)}\bigg\vert_{\mathrm{Fig}. \ref{fig:gluonreal}(d)}=&\mu^{2\epsilon}\int\frac{d^{4-2\epsilon} k}{(2\pi)^{4-2\epsilon}}\frac{i}{(P-k)\cdot n} g f_{d_1 c_1 a_1} n^{\gamma_1}\frac{-i}{(P-k)\cdot n}(-g f_{a_2 c_1 d_1} n^{\gamma_2})(i)(k\cdot n g^{\mu_2}_\mu -n^{\mu_2}P_\mu) \non\\
& \times(-i) (n^{\mu_1}P^{\mu}_{\perp}-k\cdot n g^{\mu_1
\mu}_{\perp})\frac{\delta(x-k^z/P^z)}{ 2(N^2_c-1)(P^z)^2}
\frac{-ig_{\gamma_1 \gamma_2}}{(P-k)^2-m_g^2}(-g_{\perp\mu_1
\mu_2}\delta^{a_1}_{a_2}),
\end{align}
which leads to
\begin{eqnarray}
 \tilde f_{g/g}^{(1)}(x, P^z)\bigg\vert_{\mathrm{Fig}.\ref{fig:gluonreal}(d)}=\frac{\alpha_s C_A}{2\pi}\left\{ \begin{aligned}
&\frac{x}{1-x},~~~~&x>1\\
&\frac{x}{x-1},~~~~&0<x<1\\
&\frac{x}{x-1},~~~~&x<0
\end{aligned}\right.
\end{eqnarray}
and the correction to light-cone PDF is
\begin{eqnarray}
  f_{g/g}^{(1)}(x, \mu)\bigg\vert_{\mathrm{Fig}.\ref{fig:gluonreal}(d)}=0.
\end{eqnarray}

Fig.~\ref{fig:gluonreal}(e) gives
\begin{align}
x\t f_{g/g}^{(1)}\bigg\vert_{\mathrm{Fig}.
\ref{fig:gluonreal}(e)}=&\mu^{2\epsilon}\int\frac{d^{4-2\epsilon}
k}{(2\pi)^{4-2\epsilon}}[f_{c_1 b_2 a_2}f_{c_1 b_1 a_1}(
g^{\nu_1\nu_2 }g^{\mu_1 \mu_2}-g^{\mu_2 \nu_1}g^{\mu_1
\nu_2})+f_{c_1 b_2 a_1}f_{c_1 b_1 a_2}( g^{\nu_1\nu_2 }g^{\mu_1
\mu_2}-g^{\mu_1 \nu_1}g^{\mu_2 \nu_2})]\non\\&\times(-i
g_s^2)(i)(k\cdot n g^{\nu'_2}_{\perp\mu}
-n^{\nu'_2}k_{\perp\mu})(-i)(n^{\nu'_1}k^{\mu}_{\perp} -k\cdot n
g^{\nu'_1 \mu}_{\perp}) \nonumber\\
&  \times \delta^{b_1}_{b_2} \frac{\delta(x-k^z/P^z)}{
2(N^2_c-1)(P^z)^2}\frac{-ig_{\nu_1
\nu'_1}}{k^2-m_g^2}\frac{-ig_{\nu_2
\nu'_2}}{k^2-m_g^2}(-g_{\perp\mu_1 \mu_2}\delta^{a_1}_{a_2}).
\end{align}
The amplitude gives
\begin{eqnarray}
\t f_{g/g}^{(1)}(x,
P^z)\bigg\vert_{\mathrm{Fig}.\ref{fig:gluonreal}(e)}=\frac{\alpha_s
C_A}{2\pi}\left\{
\begin{aligned}
&\frac12,~~~~&x>1\\
&\frac12,~~~~&0<x<1\\
&-\frac12,~~~~&x<0
\end{aligned}
\right.
\end{eqnarray}
and
\begin{eqnarray}
f_{g/g}^{(1)}(x,\mu)\bigg\vert_{\mathrm{Fig}.\ref{fig:gluonreal}(e)}=0.
\end{eqnarray}

\begin{figure}[htbp]
\includegraphics[width=0.55\textwidth]{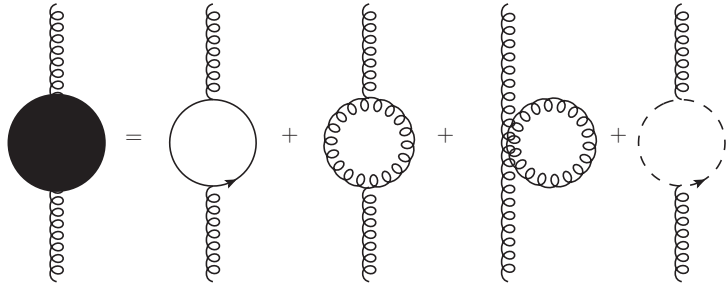}
\caption{One-loop gluon self energy}\label{fig:gluonselfenergy}
\end{figure}

Now we turn to the one-loop virtual corrections.
Fig.~\ref{fig:gluonvirtual}(a) and its conjugate diagram are just
the vacuum polarization diagrams of QCD. The black disk denotes the
four diagrams which contribute to the gluon self-energy, and these
diagrams are shown in Fig.~\ref{fig:gluonselfenergy}.
For light-cone PDF, it reads
\begin{align}
\delta
\mathcal{C}^{(1)}_{gg}\bigg\vert^{(\mathrm{subtracted})}_{\mathrm{Fig}.\ref{fig:gluonvirtual}(a)}=\frac{\alpha_s}{4\pi}\left[\left(\frac53
C_A \ln\frac{\mu^2}{m_g^2}-\frac43 T_F n_f
\ln\frac{\mu^2}{-m_g^2}\right)+\left(\frac{55}{9}-\sqrt{3}\pi\right)C_A-\frac{20}{3}T_F
  n_f\right],
\end{align}
where $n_f$ is the active quark flavor numbers. The UV divergence is
subtracted in the $\overline{\mathrm{MS}}$ scheme. For quasi PDF, it
contributes
\begin{align}
  \delta
\t
{\mathcal{C}}^{(1)}_{gg}\bigg\vert_{\mathrm{Fig}.\ref{fig:gluonvirtual}(a)}=\frac{\alpha_s}{4\pi}\left[\frac{1}{\epsilon}\left(\frac53
C_A-\frac43 T_F n_f\right)+\left(\frac53 C_A
\ln\frac{\t\mu^2}{m_g^2}-\frac43 T_F n_f
\ln\frac{\t\mu^2}{-m_g^2}\right)+\left(\frac{55}{9}-\sqrt{3}\pi\right)C_A-\frac{20}{3}T_F
  n_f\right],
\end{align}
before the UV subtraction. As we have shown in the last section, there
is no $\t\mu$ dependence but $P^z$ in the real corrections of quasi
PDFs. Therefore, we fix the scale $\t\mu$ to $P^z$ in self-energy
correction of the external gluon line and subtract the UV pole. Such
a procedure yields
\begin{align}
    \delta
\t
{\mathcal{C}}^{(1)}_{gg}\bigg\vert^{(\mathrm{subtracted})}_{\mathrm{Fig}.\ref{fig:gluonvirtual}(a)}=\frac{\alpha_s}{4\pi}\left[\left(\frac53
C_A \ln\frac{(P^z)^2}{m_g^2}-\frac43 T_F n_f
\ln\frac{(P^z)^2}{-m_g^2}\right)+\left(\frac{55}{9}-\sqrt{3}\pi\right)C_A-\frac{20}{3}T_F
  n_f\right].
\end{align}

Fig.~\eqref{fig:gluonvirtual}(b)  leads to
\begin{align}
\delta\t {\mathcal{C}}^{(1)}_{gg}\bigg\vert_{\mathrm{Fig}. \ref{fig:gluonvirtual}(b)}=&\mu^{2\epsilon}\int\frac{d^{4-2\epsilon} k}{(2\pi)^{4-2\epsilon}}(-gf_{a_1 b_1 c_1}[(P+k)^{\gamma_1} g^{\mu_1 \nu_1}+(-2k+P)^{\mu_1}g^{\nu_1 \gamma_1}+(k-2P)^{\nu_1}g^{\mu_1 \gamma_1}])\non\\
&\times (i)(n\cdot P g^{\mu_2}_{\perp\mu}
-n^{\mu_2}P_{\perp\mu})(-i)(n^{\nu'_1}k^{\mu}_{\perp} -n\cdot P
g^{\nu'_1 \mu}_{\perp})
\frac{\delta(x-1)}{ 2(N^2_c-1)(P^z)^2}\non\\
&\times  \frac{-i}{(P-k)\cdot n }(gf_{a_2 c_1 b_1}
n^{\gamma_2})\frac{-ig_{\nu_1 \nu'_1}}{k^2-m_g^2}\frac{-ig_{\gamma_1
\gamma_2}}{(P-k)^2-m_g^2}(-g_{\perp\mu_1 \mu_2}\delta^{a_1}_{a_2}).
\end{align}
Including its conjugate diagram, we have
\begin{eqnarray}
&&\delta\t
{\mathcal{C}}^{(1)}_{gg}\bigg\vert_{\mathrm{Fig}.\ref{fig:gluonvirtual}(b)}
=-\frac{\alpha_s C_A}{2\pi}\int dy \left\{
\begin{aligned}
&\frac{1+y}{1-y}\ln\frac{y}{y-1}+\frac{1-2y}{1-y},~~&y>1 \\
&\frac{1+y}{1-y}\ln\frac{4y(1-y)(P^z)^2}{(1-y+y^2)m_g^2}-\frac{2y^2-2y+1}{1-y},~~&0<y<1\\
&\frac{1+y}{1-y}\ln\frac{y-1}{y}+\frac{2y-1}{1-y},~~&y<0
\end{aligned}\right.
\end{eqnarray}
For light-cone PDF we have
\begin{eqnarray}
\delta
\mathcal{C}^{(1)}_{gg}\bigg\vert_{\mathrm{Fig}.\ref{fig:gluonvirtual}(b)}={-\frac{\alpha_s
C_A}{2\pi}\int^1_0 dy
\frac{1+y}{1-y}\ln\frac{\mu^2}{m_g^2(1-y+y^2)}}.
\end{eqnarray}

Fig.~\eqref{fig:gluonvirtual}(c)  gives
\begin{align}
\delta\t {\mathcal{C}}^{(1)}_{gg}\bigg\vert_{\mathrm{Fig}. \ref{fig:gluonvirtual}(c)}=&\mu^{2\epsilon}\int\frac{d^{4-2\epsilon} k}{(2\pi)^{4-2\epsilon}}\frac{i}{(P-k)\cdot n} g f_{d_1 c_1 a_1} n^{\gamma_1}\frac{-i}{(P-k)\cdot n}(g f_{a_2 c_2 d_1} n^{\gamma_2})(i)(n\cdot P g^{~\mu_2}_{\perp\mu} -n^{\mu_2}P_{\perp\mu})\non\\
&(-i)(n^{\mu_1}P^{\mu}_{\perp}-k\cdot n g^{\mu_1 \mu}_{\perp})
\frac{\delta(x-1)}{ 2(N^2_c-1)(P^z)^2}\frac{-ig_{\gamma_1
\gamma_2}}{(P-k)^2-m_g^2}(-g_{\perp\mu_1 \mu_2}\delta^{a_1}_{a_2}).
\end{align}
This diagram and its conjugate diagram contribute
\begin{eqnarray}
\delta \t
{\mathcal{C}}^{(1)}_{gg}\bigg\vert_{\mathrm{Fig}.\ref{fig:gluonvirtual}(c)}=-\frac{\alpha_s
C_A}{2\pi}\int dy\left\{\begin{aligned}
&\frac{y}{1-y},~~~~&y>1\\
&\frac{y}{y-1},~~~~&0<y<1\\
&\frac{y}{y-1},~~~~&y<0
\end{aligned}\right.
\end{eqnarray}
For light-cone PDF, we have vanishing results:
\begin{eqnarray}
\delta
\mathcal{C}^{(1)}_{gg}\bigg\vert_{\mathrm{Fig}.\ref{fig:gluonvirtual}(c)}=0.
\end{eqnarray}

According to Eq.~\eqref{eq:supportprops}, the contributions from the
``crossed'' diagrams can be derived by replacing $x$ with $-x$ and
multiplying an additional factor $(-1)$.

\subsection{Gluon in quark}

\begin{figure}[htbp]
\includegraphics[width=0.15\textwidth]{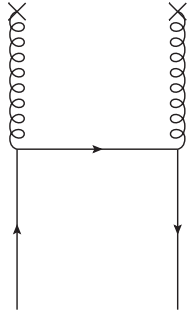}
\caption{One-loop diagrams for quark splitting to gluon.}
\label{fig:quark2gluon}
\end{figure}

The gluon in quark distribution function is shown in
Fig.~\ref{fig:quark2gluon}, which gives
\begin{align}
\t f_{g/q}(x, P^z)=& \mu^{2\epsilon}\int\frac{d^{4-2\epsilon} k}{(2\pi)^{4-2\epsilon}}\bar u(P)(-ig T^a\gamma_{\mu_2})\frac{i}{\slashed p-\slashed k} (-igT^a\gamma_{\mu_1})u(P)\non\\
&\times (i)(k\cdot n g^{\mu_2}_{\perp\mu}
-n^{\mu_2}k_{\perp\mu})(-i)(n^{\mu_1}k^{\mu}_\perp -k\cdot n g^{\mu_1
\mu}_{\perp})\frac{-i}{k^2-m_g^2}\frac{-i}{k^2-m_g^2}\frac{\delta(x-\frac{k^z}{P^z})}{
2xN_c(P^z)^2},
\end{align}

 The results for quasi PDF read
\begin{eqnarray}
\t f^{(1)}_{g/q}(x , P^z)=\frac{\alpha_s C_F}{2\pi x}\left\{
\begin{aligned}
&(1+(1-x)^2)\ln\frac{x}{x-1}-x+\frac52,~~~~&x>1\\
&(1+(1-x)^2)\ln\frac{4x (P^z)^2}{m_g^2}-2x^2+5x-\frac52,~~~~&0<x<1\\
&(1+(1-x)^2)\ln\frac{x-1}{x}+x-\frac52,~~~~&x<0
\end{aligned}\right.
\end{eqnarray}
while for light-cone  one, we have
\begin{eqnarray}
f^{(1)}_{g/q}(x, \mu)=\frac{\alpha_s C_F}{2\pi x}\left\{
\begin{aligned}
&0,~~~~&x>1,~~x<0\\
&\left(1+(1-x)^2\right)\ln\frac{\mu^2}{m_g^2(1-x)}-2(x^2-x+1),~~&0<x<1
\end{aligned}
\right.
\end{eqnarray}

The contributions from the ``crossed'' diagrams can be derived by
using Eq.~\eqref{eq:supportprops}.

\subsection{Quark in gluon}

\begin{figure}[htbp]
\includegraphics[width=0.15\textwidth]{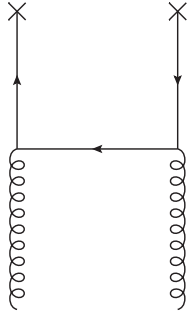}
\caption{One-loop diagrams for gluon splitting to quark.}
\label{fig:gluon2quark}
\end{figure}

The quark in gluon distribution function is shown in
Fig.~\ref{fig:gluon2quark},
which contributes to
\begin{align}
\t f^{(1)}_{q/g}(x, P^z)=& \mu^{2\epsilon}\int\frac{d^{4-2\epsilon}
k}{(2\pi)^{4-2\epsilon}}(-1)\mathrm{Tr}\left[\gamma^z\frac{i}{\slashed k}
(-ig T^a\gamma^{\mu})\frac{i}{\slashed k-\slashed p}
(-igT^a\gamma^{\nu})\frac{i}{\slashed
k}\right](-g_{\perp\mu\nu})\frac{\delta(x-\frac{k^z}{P^z})}{
4(N_c^2-1)P^z},
\end{align}
The result reads
\begin{eqnarray}
\widetilde f_{q/g}(x,P^z)=\frac{\alpha_s T_F}{2\pi}\left\{
\begin{aligned}
&(x^2+(1-x)^2)\ln\frac{x}{x-1}-2x+1,~~~~&x>1\\
&(x^2+(1-x)^2)\ln\frac{4(P^z)^2}{-m_g^2}-4x^2+2x,~~~~&0<x<1\\
&(x^2+(1-x)^2)\ln\frac{x-1}{x}+2x-1,~~~~&x<0
\end{aligned}\right.
\end{eqnarray}
For light-cone distribution, the result is
\begin{eqnarray}
f^{(1)}_{q/g}(x,\mu)=\frac{\alpha_s T_F}{2\pi}\left\{
\begin{aligned}
&0,~~~~&x>1,~~x<0\\
&(x^2+(1-x)^2)\ln\frac{\mu^2}{-m_g^2 x(1-x)}-1,~~~~&0<x<1
\end{aligned}
\right.
\end{eqnarray}

The contributions from ``crossed'' diagrams can be derived by using
Eq.~\eqref{eq:supportprops}.

\subsection{Matching}

In this subsection we perform the matching between the
quasi and light-cone PDFs. According to LaMET, if the factorization
theorem holds at the leading power of $\mu/P^z$,   we   expect a matching equation as
\begin{align}
  \t f_{i/H}(x, P^z)=\int^{1}_{0}\frac{dy}{y}
  Z_{ij}\left(\frac{x}{y},\frac{\mu}{P^z}\right)f_{j/H}(y, \mu)\equiv Z_{ij}\left(\xi, \frac{\mu}{P^z}\right)\otimes
  f_{j/H}(y),\label{eq:matchingeq}
\end{align}
where $\xi\equiv x/y$, and $i, j$ denotes the involved parton such as quark $q$ and gluon $g$. To perform the perturbative matching, one should first
replace the hadron with a parton $k$. Then $\t f_{i/k}(x, P^z)$,
$f_{j/k}(y, \mu)$ and the matching function $Z_{ij}(\xi, \mu/P^z)$
can be calculated in perturbation theory. They can be expanded as
series of $\alpha_s$
\begin{align}
  \t f_{i/k}(x, P^z)&=\sum_{n=0}^{\infty} \t f^{(n)}_{i/k}(x, P^z),\nonumber\\
f_{j/k}(y,\mu)&=\sum_{n=0}^{\infty} f^{(n)}_{j/k}(y, \mu) ,\nonumber\\
Z_{ij}\left(\xi,\frac{\mu}{P^z}\right)&=\sum_{n=0}^{\infty}\left(\frac{\alpha_s}{2\pi}\right)^n
Z^{(n)}_{ij}\left(\xi, \frac{\mu}{P^z}\right).\label{eq:coeffs}
\end{align}
where
\begin{align}
  f^{(0)}_{j/k}(y)=\delta_{jk}\delta(1-y),
  ~~~~\t f^{(0)}_{i/k}(x)=\delta_{ik}\delta(1-x)
\end{align}
are the tree-level distributions. At the order
$\mathcal{O}(\alpha^0_s)$, we have
$Z^{(0)}_{ij}(\xi,\mu/P^z)=\delta_{ij}\delta(1-\xi)$.

Then we calculate the matching coefficient $Z^{(1)}_{ij}$. At the
order of $\mathcal{O}(\alpha^1_s)$, the matching equation
Eq.~\eqref{eq:matchingeq} is reduced to
\begin{align}
\t f^{(1)}_{i/k}(x, P^z)&=
Z^{(0)}_{ij}\left(\xi,\frac{\mu}{P^z}\right)\otimes
f^{(1)}_{j/k}(y,\mu)+\frac{\alpha_s}{2\pi}
Z^{(1)}_{ij}\left(\xi,\frac{\mu}{P^z}\right)\otimes
f^{(0)}_{j/k}(y)\nonumber\\
&=
f^{(1)}_{i/k}(x,\mu)+\frac{\alpha_s}{2\pi}Z^{(1)}_{ik}\left(x,\frac{\mu}{P^z}\right),
\end{align}
and then we obtain
\begin{align}
 \frac{\alpha_s}{2\pi} Z^{(1)}_{ik}\left(x,\frac{\mu}{P^z}\right)=\t f^{(1)}_{i/k}(x,
  P^z)-f^{(1)}_{i/k}(x, \mu).\label{eq:match:oneloop}
\end{align}

Eq.~\eqref{eq:match:oneloop} implies  that, the one loop correction to
the matching function $Z_{ik}(\xi,\mu/P^z)$, can be derived by
taking the difference of the one-loop quasi and LC PDFs. With the
one-loop corrections calculated in Sec.~\ref{sec:1loop}, we have
\begin{align}
   Z^{(1)}_{qq}\left(\xi,\frac{\mu}{P^z}\right)&=C_F\left\{\begin{aligned}
    &-\frac{1+\xi^2}{1-\xi}\ln\frac{\xi-1}{\xi}+1,   ~~~~&\xi>1\\
    &-\frac{1+\xi^2}{1-\xi}\ln\frac{\mu^2}{4(P^z)^2\xi(1-\xi)}+\frac{2-5\xi+\xi^2}{1-\xi},~~~~&0<\xi<1\\
&\frac{1+\xi^2}{1-\xi}\ln\frac{\xi-1}{\xi}-1,   ~~~~&\xi<0
  \end{aligned}\right.\nonumber\\
  &+\delta(1-\xi)C_F\int d\eta \left\{\begin{aligned}
    &\frac{1+\eta^2}{1-\eta}\ln\frac{\eta-1}{\eta}-1,~~~~&\eta>1\\
    &\frac{1+\eta^2}{1-\eta}\ln\frac{\mu^2}{4(P^z)^2\eta(1-\eta)}-\frac{2-5\eta+\eta^2}{1-\eta},~~~~&0<\eta<1\\
&-\frac{1+\eta^2}{1-\eta}\ln\frac{\eta-1}{\eta}+1,   ~~~~&\eta<0
  \end{aligned}\right.\nonumber\\
  &=C_F\left\{\begin{aligned}
    &\left[-\frac{1+\xi^2}{1-\xi}\ln\frac{\xi-1}{\xi}+1\right]_+,   ~~~~&\xi>1\\
    &\left[-\frac{1+\xi^2}{1-\xi}\ln\frac{\mu^2}{4(P^z)^2\xi(1-\xi)}+\frac{2-5\xi+\xi^2}{1-\xi}\right]_+,~~~~&0<\xi<1\\
&\left[\frac{1+\xi^2}{1-\xi}\ln\frac{\xi-1}{\xi}-1\right]_+,
~~~~&\xi<0
\end{aligned}\right.
\end{align}
where $[\cdots]_+$ denotes the ``plus'' distribution
\begin{align}
  \int \left[f(x)\right]_+T(x)dx\equiv \int
  f(x)\left[T(x)-T(1)\right]dx.
\end{align}
The above one loop matching factor for quark to quark is equivalent to the result of Ref.~\cite{Xiong:2013bka}. Note that there is a little difference if one directly use
the result of Ref.~\cite{Xiong:2013bka}, since therein a cut-off scheme is employed to regularize the UV divergence while a small quark mass is employed to regularize the collinear divergence.

The quark to gluon and    gluon to quark ones  are given as
\begin{align}
   Z^{(1)}_{qg}\left(\xi,\frac{\mu}{P^z}\right)&=T_F\left\{\begin{aligned}
    &(\xi^2+(1-\xi)^2)\ln\frac{\xi}{\xi-1}-2\xi+1,   ~~~~&\xi>1\\
    &-(\xi^2+(1-\xi)^2)\ln\frac{\mu^2}{4\xi(1-\xi)(P^z)^2}+1+2\xi-4\xi^2,~~~~&0<\xi<1\\
  &-(\xi^2+(1-\xi)^2)\ln\frac{\xi}{\xi-1}+2\xi-1,   ~~~~&\xi<0\\
  \end{aligned}\right.
\end{align}
and
\begin{align}
   Z^{(1)}_{gq}\left(\xi,\frac{\mu}{P^z}\right)&=C_F\left\{\begin{aligned}
    &\frac{1+(1-\xi)^2}{\xi}\ln\frac{\xi}{\xi-1}-1+\frac{5}{2\xi},   ~~~~&\xi>1\\
    &-\frac{1+(1-\xi)^2}{\xi}\ln\frac{\mu^2}{4\xi(1-\xi)(P^z)^2}+3-\frac{1}{2\xi},~~~~&0<\xi<1\\
   &-\frac{1+(1-\xi)^2}{\xi}\ln\frac{\xi}{\xi-1}+1-\frac{5}{2\xi},   ~~~~&\xi<0\\
  \end{aligned}\right.
\end{align}

The gluon to gluon matching function is given as
\begin{align}
   &~~~~Z^{(1)}_{gg}\left(\xi,\frac{\mu}{P^z}\right)\nonumber\\
   &=C_A\left\{\begin{aligned}
    &\left(\frac{2\xi^3-3\xi^2+2\xi-2}{\xi}+\frac{\xi(1+\xi)}{\xi-1}\right)\ln\frac{\xi-1}{\xi}+2\xi-1+\frac{8}{3\xi},   ~~~~&\xi>1\\
    &\left(\frac{2\xi^3-3\xi^2+2\xi-2}{\xi}+\frac{\xi(1+\xi)}{\xi-1}\right)\ln\frac{\mu^2}{4\xi(1-\xi)(P^z)^2}\\
    &+\frac{10\xi^2}{3}-4\xi+4-\frac{2}{3\xi}-\frac{2\xi^2-\xi+1}{1-\xi},~~~~&0<\xi<1\\
   &-\left(\frac{2\xi^3-3\xi^2+2\xi-2}{\xi}+\frac{\xi(1+\xi)}{\xi-1}\right)\ln\frac{\xi-1}{\xi}-2\xi+1-\frac{8}{3\xi},   ~~~~&\xi<0\\
  \end{aligned}\right.\nonumber\\
  &~~~~-\delta(1-\xi)C_A\int d\eta \left\{
  \begin{aligned}
  &\frac{1+\eta}{1-\eta}\ln\frac{\eta}{\eta-1}+1,~~~~&\eta>1\\
    &\frac{1+\eta}{1-\eta}\ln\frac{4\eta(1-\eta)(P^z)^2}{\mu^2}-\left(\frac53 C_A-\frac43 T_F n_f\right)\ln\frac{(P^z)^2}{\mu^2}\\
    &-\frac{2\eta^2-\eta+1}{1-\eta},~~~~&0<\eta<1\\
&-\frac{1+\eta}{1-\eta}\ln\frac{\eta}{\eta-1}-1,~~~~&\eta<0\\
  \end{aligned}\right.\nonumber\\
  &=C_A\left\{\begin{aligned}
    &\frac{2\xi^3-3\xi^2+2\xi-2}{\xi}\ln\frac{\xi-1}{\xi}+\xi\left[\frac{1+\xi}{\xi-1}\ln\frac{\xi-1}{\xi}+1\right]_+ +\xi-1+\frac{8}{3\xi},   ~~~~&\xi>1\\
    &\frac{2\xi^3-3\xi^2+2\xi-2}{\xi}\ln\frac{\mu^2}{4\xi(1-\xi)(P^z)^2}+\xi\left[\frac{1+\xi}{\xi-1}\ln\frac{\mu^2}{4\xi(1-\xi)(P^z)^2}\right]_+\\
    &+\delta(1-\xi)\left(\frac53 C_A-\frac43 T_F
  n_f\right)\ln\frac{(P^z)^2}{\mu^2} -\left[\frac{2\xi^2-\xi+1}{1-\xi}\right]_+ +\frac{10\xi^2}{3}-4\xi+4-\frac{2}{3\xi},~~~~&0<\xi<1\\
   &-\frac{2\xi^3-3\xi^2+2\xi-2}{\xi}\ln\frac{\xi-1}{\xi}-\xi\left[\frac{1+\xi}{\xi-1}\ln\frac{\xi-1}{\xi}+1\right]_+ -\xi+1-\frac{8}{3\xi},   ~~~~&\xi<0\\
  \end{aligned}\right.
\end{align}

Now we consider the contribution from the ``crossed'' diagrams. As
we have discussed in Sec.~\ref{sec:1loop}, ``crossed'' diagrams give
non-zero contribution to light-cone distributions in $-1<x<0$, and
to quasi distributions in $x>0$, $-1<x<0$ and $x<-1$. The
contribution from the ``uncrossed'' and ``crossed'' diagrams can be
related by Eq.~\eqref{eq:supportprops}. For quark, the ``crossed''
diagrams can be explained as corrections to anti-quark
distributions. By replacing $x$ with $-x$ and $y$ with $-y$ in
Eq.~\eqref{eq:matchingeq}, one can immediately finds that the
corrections from ``crossed'' diagrams share the same matching
function $Z(x/y, \mu/P^z)$. Then, the matching equation
Eq.~\eqref{eq:matchingeq} can be extended to
\begin{align}
  \t f_{i/H}(x, P^z)=\int^{1}_{-1}\frac{dy}{|y|}
  Z_{ij}\left(\frac{x}{y},\frac{\mu}{P^z}\right)f_{j/H}(y, \mu).\label{eq:matchingeq:full}
\end{align}
In this equation, the light-cone distributions have non-zero support
in $-1<x<1$. The anti-quark distributions are also included.

\subsection{$P^z$ evolution equations}

From the matching equation Eq.~\eqref{eq:matchingeq} we can derive
the evolution equations with $P^z$. Notice that the light-cone PDFs
is independent of $P^z$, one can take the derivative of left
and right hand sides of Eq.~\eqref{eq:matchingeq} with $\ln P^z$,
which yields
\begin{align}
  \frac{d \t f_{i/H}(x, P^z)}{d\ln P^z}&=\frac{d Z_{ik}\left(\frac{x}{z},\frac{\mu}{P^z}\right)}{d \ln
  P^z}\otimes f_{k/H}(z,\mu)\nonumber\\
  &=\frac{d Z_{ik}\left(\frac{x}{z},\frac{\mu}{P^z}\right)}{d \ln
  P^z}\otimes
  \left[Z^{-1}\left(\frac{z}{y},\frac{\mu}{P^z}\right)\right]_{kj}\otimes
  \t f_{j/H}(y, P^z)\nonumber\\
  &=\frac{d\ln Z_{ij}\left(\frac{x}{y},\frac{\mu}{P^z}\right)}{d\ln
  P^z}\otimes\t f_{j/H}(y, P^z)\nonumber\\
  &=\t P_{i\leftarrow j}(\xi) \otimes\t f_{j/H}(y, P^z),
\end{align}
where $\t P_{i\leftarrow j}(\xi)\equiv d\ln Z_{ij}(x/y,\mu/P^z)/d\ln
P^z$ is the evolution kernel, and $\xi\equiv x/y$. With the matching
functions calculated in the last subsection, we obtain the evolution
equations at  one loop level as follows:
\begin{eqnarray}
  \frac{ d \tilde f_{q/H}(x,  P^z) }{d \ln P^z}& = &\frac{\alpha_s}{\pi}\int \frac{dy}{y} \left[\tilde P_{q\leftarrow q}\left(\frac{x}{y}\right)  \tilde f_{q/H}\left(y,\frac{\mu}{P^z}\right) +\tilde P_{q\leftarrow g}\left(\frac{x}{y}\right) \tilde f_{g/H}\left(y,\frac{\mu}{P^z}\right)\right]\ ,\\
    \frac{ d \tilde f_{g/H}(x,  P^z)}{d \ln P^z} & = &\frac{\alpha_s}{\pi}\int \frac{dy}{y} \left[\tilde P_{g\leftarrow q}\left(\frac{x}{y}\right)  \sum_f \left(\tilde f_{q_f/H}\left(y,\frac{\mu}{P^z}\right) +\tilde{f}_{\bar{q}_f/H}\left(y,\frac{\mu}{P^z}\right)\right)+\tilde P_{g\leftarrow g}\left(\frac{x}{y}\right)  \tilde f_{g/H}\left(y,\frac{\mu}{P^z}\right)\right],
\end{eqnarray}
where the evolution  kernels are
\begin{eqnarray}
  \tilde P_{q\leftarrow q}(\xi)  & = &C_F\left\{ \begin{array} {ll} 0\ ,\hspace{2.5cm} \xi>1\  \mbox{or}\  \xi<0
  \\ \frac{1+\xi^2}{(1-\xi)^+}+\frac{3}{2}\delta(1-\xi)\ , \hspace{0.5cm}0<\xi\leq1\end{array} \right.\\
   \tilde P_{q\leftarrow g}(\xi)  & = &T_F\left\{ \begin{array} {ll} 0\ ,& \xi>1\  \mbox{or}\  \xi<0
   \\ \xi^2+(1-\xi)^2\ , & 0<\xi\leq1 \end{array} \right.\\
    \tilde P_{g\leftarrow q}(\xi)  & = &C_F\left\{ \begin{array} {ll} 0\ ,& \xi>1\  \mbox{or}\  \xi<0
    \\ \frac{1+(1-\xi)^2}{\xi}\ , & 0<\xi\leq1 \end{array} \right.\\
     \tilde P_{g\leftarrow g}(\xi)  & = &\left\{ \begin{array} {ll} 0\ ,\hspace{2.5cm}  \xi>1\  \mbox{or}\  \xi<0
     \\ \frac{2C_A(1-\xi+\xi^2)^2}{\xi(1-\xi)^+}+\frac{\beta_0}{2}\delta(1-\xi)\ ,   0<\xi\leq1 \end{array} \right.
\end{eqnarray}
Here  $\beta_0=({11C_A-2n_f})/{3}$. The  $P^z$ evolution equations of
quasi distributions    are just the DGLAP equations for the
light-cone parton distribution functions~\cite{Altarelli:1977zs}.

\section{Conclusions}

In this work, we have  investigated   the unpolarized gluon quasidistribution function  in the
nucleon at one loop level in the large momentum effective theory.  To regularize  the ultraviolet divergences, we have adopted the cut-off scheme and the dimensional regularization scheme, while two schemes with  finite gluon mass or the offshellness are used to regulate the infrared divergences.   In addition to the ordinary quark and gluon distribution functions, we  have also studied the quark to gluon and gluon to quark contributions.

When studying  the quark quasidistribution
in the cut-off scheme,  one can find that  power law ultraviolet divergences arise  in
the nonlocal operator and they all are subjected  to the Wilson lines. On the contrary  in the gluon quasidistribution,  we have pointed out in this work  that the linear  ultraviolet
divergences also exist in  the real diagram without connecting to the Wilson
line.  The one loop matching
factors between the parton quasi and light cone distribution functions  have been  derived. Through the simulation on the lattice of the parton quasidistributions, one can finally obtain the nonperturbative information of the parton distributions.  At last, we have  studied  the $P^z$ evolution equation for the quasi  parton
distribution functions, and   found that   the $P^z$ evolution kernels are identical to the DGLAP kernels.

\section*{Acknowledgments}

We  are grateful to Jun Gao,   Tomomi Ishikawa, Xiangdong Ji, Yu
Jia, Hsiang-Nan Li, Xiaohui Liu, Yanqing Ma,   Fan Wang, Feng Yuan, Jianhui
Zhang and  Yong Zhao  for helpful discussions. This work was
supported in part by the National Natural Science Foundation of
China under Grant No. 11647163, 11575110, 11655002, by Natural
Science Foundation of Shanghai under Grant No.~15DZ2272100 and
No.~15ZR1423100,  by Natural Science Foundation of Jiangsu under
Grant No.~BK20171471, by the Young Thousand Talents Plan,   by Key
Laboratory for Particle Physics, Astrophysics and Cosmology,
Ministry of Education, and by the Research Start-up Funding of
Nanjing Normal University.

\appendix

\section{Cut-off scheme results}
\label{sec:appendix_cutoff}

In this appendix, we will present the one-loop results for the parton quasi (light cone) distribution functions at the cut-off scheme.

We first take Fig.~\ref{fig:gluonreal}(a) as an example to illustrate  the  calculation in UV cut-off scheme. By
contracting the Lorentz and color indices for
Eq.~\eqref{eq:gluonreal:amp1}, and integrating out the delta
function at the non-local vertex, we have the contribution
\begin{align}
  &~~~~\t f_{g/g}(x, P^z)\bigg\vert_{\mathrm{Fig}.\ref{fig:gluonreal}(a)}\nonumber\\
  &=-\frac{2i\alpha_s C_A}{x P^z}\int\frac{d k^0
  d^{2}k_{\perp}}{(2\pi)^3}\frac{1}{\left(-(k^0)^2+k^2_{\perp}+m_g^2+x^2(P^z)^2\right)^2\left(-(k^0)^2+2k^0\sqrt{m_g^2+(P^z)^2}+k^2_{\perp}+(P^z)^2x(x-2)\right)}\nonumber\\
&\times\bigg[x^2 (P^z)^2
\left(-(k^0)^2-2k^0\sqrt{m_g^2+(P^z)^2}-m_g^2+(P^z)^2x(x+2)\right)+2(k^2_{\perp})^2+k^2_{\perp}
  (P^z)^2(5x^2+4)\bigg].\label{eq:app:cutoff}
\end{align}
The two dominators are given as
\begin{align}
  D_1&\equiv -(k^0)^2+k^2_{\perp}+m_g^2+x^2(P^z)^2,\nonumber\\
D_2&\equiv
-(k^0)^2+2k^0\sqrt{m_g^2+(P^z)^2}+k^2_{\perp}+(P^z)^2x(x-2).
\end{align}
The integral in Eq.~\eqref{eq:app:cutoff} can be reduced to five scalar integrals:
\begin{align}
  I_1&=\int\frac{d k^0
  d^{2}k_{\perp}}{(2\pi)^{3}}\frac{(k^0)^4}{D^2_1
  D_2},\nonumber\\
   I_2&=\int\frac{d k^0
  d^{2}k_{\perp}}{(2\pi)^{3}}\frac{k^0}{D^2_1
  D_2},\nonumber\\
   I_3&=\int\frac{d k^0
  d^{2}k_{\perp}}{(2\pi)^{3}}\frac{1}{D^2_1
  D_2},\nonumber\\
     I_4&=\int\frac{d k^0
  d^{2}k_{\perp}}{(2\pi)^{3}}\frac{(k^2_{\perp})^2}{D^2_1
  D_2},\nonumber\\
     I_5&=\int\frac{d k^0
  d^{2}k_{\perp}}{(2\pi)^{3}}\frac{k^2_{\perp}}{D^2_1
  D_2}.
\end{align}

To calculate these integrals, we   perform the Feynman
parametrization,  and then integrate over $k^0$ by using the residue theorem. The remaining integral is the integral over transverse momentum $k_{\perp}$. The possible UV divergence is regularized by
introducing a cut off $\Lambda$ on the transverse momentum. The last step is to integrate over the Feynman parameter. Through  the above procedure,
we arrive at
\begin{align}
 I_1&=-\frac{i P^z}{4\pi}\left\{ \begin{aligned}
&x^3\ln\frac{x-1}{x}+x^2+\frac{x}{8}+\frac{1}{8(x-1)}+\frac{11}{24}+\frac38 \frac{\Lambda}{P^z},~~~~&x>1\\
&x^3 \ln\frac{(1-x+x^2)m^2_g}{4x(1-x)(P^z)^2}-\frac{(x-1) x^4 (P^{z})^2}{2 \left(1-x+x^2\right) m_g^2}\\
&+\frac{64 x^8-216 x^7+374 x^6-394 x^5+252 x^4-110 x^3+43 x^2-24 x+8}{24 (x-1) (1-x+x^2)^2}+\frac38 \frac{\Lambda}{P^z},   &0<x<1\\
&x^3\ln\frac{x}{x-1}-x^2-\frac{x}{8}-\frac{1}{8(x-1)}-\frac{11}{24}+\frac38
\frac{\Lambda}{P^z},~~~~&x<0
\end{aligned}\right.
\\
 I_2&=-\frac{i}{16\pi (P^z)^2 }\left\{ \begin{aligned}
&\ln\frac{x-1}{x}+\frac{2x-1}{2x(x-1)},~~~~&x>1\\
&\ln\frac{4x(1-x)(P^z)^2}{(1-x+x^2)m^2_g}-\frac{2(x-1)x (P^z)^2}{(1-x+x^2)m^2_g}+\frac{2x^6-6x^5+10x^4-10x^3+2x^2+2x-1}{2x(x-1)(1-x+x^2)^2},   &0<x<1\\
&-\ln\frac{x-1}{x}-\frac{2x-1}{2x(x-1)},~~~~&x<0
\end{aligned}\right.
\\
 I_3&=-\frac{i}{32\pi (P^z)^3 }\left\{ \begin{aligned}
&\frac{1}{x^2(x-1)},~~~~&x>1\\
&-\frac{4 (x-1)(P^z)^2}{m^2_g (1-x+x^2)}+\frac{2 x^4-6 x^3+4 x^2-2 x+1}{(x-1) x^2 \left(1-x+x^2\right)^2},   &0<x<1\\
&-\frac{1}{x^2(x-1)},~~~~&x<0
\end{aligned}\right.
\\
 I_4&=\frac{i P^z}{12\pi }\left\{ \begin{aligned}
&3x-1-\frac98\frac{\Lambda}{P^z},~~~~&x>1\\
&2x^3-3x^2+3x-1-\frac98\frac{\Lambda}{P^z},   &0<x<1\\
&-(3x-1)-\frac98\frac{\Lambda}{P^z},~~~~&x<0
\end{aligned}\right.
\\
 I_5&=-\frac{i }{8\pi P^z}\left\{ \begin{aligned}
&(x-1)\ln\frac{x-1}{x}+1,~~~~&x>1\\
&(x-1)\ln\frac{(1-x+x^2)m^2_g}{4x(1-x)(P^z)^2}+2x-1,   &0<x<1\\
&-(x-1)\ln\frac{x-1}{x}-1.~~~~&x<0
\end{aligned}\right.
\end{align}
One can find  that  the integrals $I_{2,3,5}$ are  UV finite, but  $I_1$ and
$I_4$ are  linearly divergent. With these integrals, we
finally arrive at Eq.~\eqref{eq:quasi:cutoff}, which are linearly divergent as well.

The results for Fig.~\ref{fig:gluonreal}(b) and
Fig.~\ref{fig:gluonreal}(c) are given as
\begin{eqnarray}
 \tilde f_{g/g}^{(1)}(x, P^z,\Lambda)\bigg\vert_{\mathrm{Fig}.\ref{fig:gluonreal}(b,c)}=-\frac{\alpha_s C_A}{4\pi}\left\{
\begin{aligned}
&\frac{1}{x-1}\left[x(1+x)\ln\frac{x}{x-1}-2x+1+\frac{\Lambda}{P^z}\right],~~~~&x>1\\
&\frac{1}{x-1}\left[x(1+x)\ln\frac{4x(1-x)(P^z)^2}{(1-x+x^2)m_g^2}-(2x^2-2x+1)+\frac{\Lambda}{P^z}\right],~~~~&0<x<1\\
&\frac{1}{x-1}\left[x(1+x)\ln\frac{x-1}{x}+2x-1+\frac{\Lambda}{P^z}\right],~~~~&x<0
\end{aligned}\right.
\end{eqnarray}
and
\begin{eqnarray}
f_{g/g}^{(1)}(x,
\Lambda)\bigg\vert_{\mathrm{Fig}.\ref{fig:gluonreal}(b,c)}=-\frac{\alpha_s
C_A}{4\pi}\left\{
\begin{aligned}
&0,~~~~&x>1,~x<0\\
&{\frac{x(1+x)}{x-1}\ln\frac{\Lambda^2}{m_g^2(1-x+x^2)}},~~~~&0<x<1
\end{aligned}
\right.
\end{eqnarray}

Fig.~\ref{fig:gluonreal}(d) gives
\begin{eqnarray}
 \tilde f_{g/g}^{(1)}(x, P^z,\Lambda)\bigg\vert_{\mathrm{Fig}.\ref{fig:gluonreal}(d)}=\frac{\alpha_s C_A}{2\pi}\left\{ \begin{aligned}
&\frac{x}{1-x}+\frac{x}{(x-1)^2}\frac{\Lambda}{P^z},~~~~&x>1\\
&\frac{x}{x-1}+\frac{x}{(x-1)^2}\frac{\Lambda}{P^z},~~~~&0<x<1\\
&\frac{x}{x-1}+\frac{x}{(x-1)^2}\frac{\Lambda}{P^z},~~~~&x<0
\end{aligned}\right.
\end{eqnarray}
and
\begin{eqnarray}
  f_{g/g}^{(1)}(x, \Lambda)\bigg\vert_{\mathrm{Fig}.\ref{fig:gluonreal}(d)}=0.
\end{eqnarray}

Fig.~\ref{fig:gluonreal}(e) gives
\begin{eqnarray}
\t f_{g/g}^{(1)}(x,
P^z,\Lambda)\bigg\vert_{\mathrm{Fig}.\ref{fig:gluonreal}(e)}=\frac{\alpha_s
C_A}{2\pi}\left\{
\begin{aligned}
&\frac12-\frac{\Lambda}{2x P^z},~~~~&x>1\\
&\frac12-\frac{\Lambda}{2x P^z},~~~~&0<x<1  \\
&-\frac12-\frac{\Lambda}{2x P^z},~~~~&x<0
\end{aligned}
\right.
\end{eqnarray}
and
\begin{eqnarray}
f_{g/g}^{(1)}(x,\Lambda)\bigg\vert_{\mathrm{Fig}.\ref{fig:gluonreal}(e)}=0.
\end{eqnarray}

For the virtual corrections, Fig.~\eqref{fig:gluonvirtual}(b)
contributes
\begin{eqnarray}
&&\delta\t
{\mathcal{C}}^{(1)}_{gg}\bigg\vert_{\mathrm{Fig}.\ref{fig:gluonvirtual}(b)}
=-\frac{\alpha_s C_A}{2\pi}\int dy \left\{
\begin{aligned}
&\frac{1+y}{1-y}\ln\frac{y}{y-1}+\frac{1-2y}{1-y}+\frac{1}{1-y}\frac{\Lambda}{P^z},~~&y>1 \\
&\frac{1+y}{1-y}\ln\frac{4y(1-y)(P^z)^2}{(1-y+y^2)m_g^2}-\frac{2y^2-2y+1}{1-y}+\frac{1}{1-y}\frac{\Lambda}{P^z},~~&0<y<1\\
&\frac{1+y}{1-y}\ln\frac{y-1}{y}+\frac{2y-1}{1-y}+\frac{1}{1-y}\frac{\Lambda}{P^z}.~~&y<0
\end{aligned}\right.
\end{eqnarray}
to quasi gluon distribution.
 For light-cone PDF we have
\begin{eqnarray}
\delta
\mathcal{C}^{(1)}_{gg}\bigg\vert_{\mathrm{Fig}.\ref{fig:gluonvirtual}(b)}={-\frac{\alpha_s
C_A}{2\pi}\int^1_0 dy
\frac{1+y}{1-y}\ln\frac{\mu^2}{m_g^2(1-y+y^2)}}.
\end{eqnarray}

Fig.~\ref{fig:gluonvirtual}(c) and its conjugate diagram contribute
\begin{eqnarray}
\delta \t
{\mathcal{C}}^{(1)}_{gg}\bigg\vert_{\mathrm{Fig}.\ref{fig:gluonvirtual}(c)}=-\frac{\alpha_s
C_A}{2\pi}\int dy\left\{\begin{aligned}
&\frac{y}{1-y}+\frac{y}{(1-y)^2}\frac{\Lambda}{P^z},~~~~&y>1\\
&\frac{y}{y-1}+\frac{y}{(1-y)^2}\frac{\Lambda}{P^z},~~~~&0<y<1\\
&\frac{y}{y-1}+\frac{y}{(1-y)^2}\frac{\Lambda}{P^z},~~~~&y<0
\end{aligned}\right.
\end{eqnarray}
to quasi gluon distribution. For light-cone PDF, these diagrams
vanish since $n_+^2=0$:
\begin{eqnarray}
\delta
\mathcal{C}^{(1)}_{gg}\bigg\vert_{\mathrm{Fig}.\ref{fig:gluonvirtual}(c)}=0
\end{eqnarray}

At last we discuss Fig.~\ref{fig:gluonvirtual}(a), the vacuum
polarization diagram of QCD. Due to the Ward identity, the
one-loop two-point green function of gluon should be transversely
polarized. However, it is well known that a cut-off on the momentum
integral breaks the gauge invariance.  When the cut-off is imposed on  all of
the four components of the loop integral, it will {lead} to a
quadratic divergence;  when the cut-off is imposed on the transverse
momentum integral, which is the case of this work ({integrating} out
``0'' component by using residue theorem and keeping $z$ component
unintegrated), it will {lead} to a linear divergence. To evaluate
Fig.~\ref{fig:gluonvirtual}(a) correctly, a cut-off scheme
respecting the gauge invariance is needed, instead of a naive
cut-off. This implies  the linear divergence in the real
corrections, as shown in the above, is a consequence of breaking the
gauge invariance in the cut-off. For this reason, we can  not have  reasonable   results for Fig.~\ref{fig:gluonvirtual}(a) at present.

\section{Integrals in dimensional regularization}
\label{sec:appendix_dr}

In dimensional regularization scheme, the space-time dimension is
shifted from $d=4$ to $d=4-2\epsilon$. However, as we will see
below, the real corrections to quasi PDFs is finite in DR scheme.

Again we will  briefly demonstrate  the calculation method by taking
Fig.~\ref{fig:gluonreal}(a) as an example. By contracting the
Lorentz indices for Eq.~\eqref{eq:gluonreal:amp1}, and integrating
out the delta function at the non-local vertex, we have
\begin{align}
  &~~~~\t f_{g/g}(x, P^z)\bigg\vert_{\mathrm{Fig}.\ref{fig:gluonreal}(a)}\nonumber\\
  &=-\frac{2i\alpha_s C_A \mu^{2\epsilon}}{(d-2)x P^z}\int\frac{d k^0
  d^{2-2\epsilon}k_{\perp}}{(2\pi)^{3-2\epsilon}}\frac{1}{\left(-(k^0)^2+k^2_{\perp}+m_g^2+x^2(P^z)^2\right)^2\left(-(k^0)^2+2k^0\sqrt{m_g^2+(P^z)^2}+k^2_{\perp}+(P^z)^2x(x-2)\right)}\nonumber\\
&\times\bigg[(d-2)x^2 (P^z)^2
\left(-(k^0)^2-2k^0\sqrt{m_g^2+(P^z)^2}-m_g^2+(P^z)^2x(x+2)\right)+4(k^2_{\perp})^2+(d-2)k^2_{\perp}
  (P^z)^2(5x^2+4)\bigg].\label{eq:app:dimreg}
\end{align}
We denote the two dominators as
\begin{align}
  D_1&\equiv -(k^0)^2+k^2_{\perp}+m_g^2+x^2(P^z)^2,\nonumber\\
D_2&\equiv
-(k^0)^2+2k^0\sqrt{m_g^2+(P^z)^2}+k^2_{\perp}+(P^z)^2x(x-2).
\end{align}
The integral can be reduced to five scalar integrals, which are
\begin{align}
  I_1&=\int\frac{d k^0
  d^{2-2\epsilon}k_{\perp}}{(2\pi)^{3-2\epsilon}}\frac{(k^0)^4}{D^2_1
  D_2},\nonumber\\
   I_2&=\int\frac{d k^0
  d^{2-2\epsilon}k_{\perp}}{(2\pi)^{3-2\epsilon}}\frac{k^0}{D^2_1
  D_2},\nonumber\\
   I_3&=\int\frac{d k^0
  d^{2-2\epsilon}k_{\perp}}{(2\pi)^{3-2\epsilon}}\frac{1}{D^2_1
  D_2},\nonumber\\
     I_4&=\int\frac{d k^0
  d^{2-2\epsilon}k_{\perp}}{(2\pi)^{3-2\epsilon}}\frac{(k^2_{\perp})^2}{D^2_1
  D_2},\nonumber\\
     I_5&=\int\frac{d k^0
  d^{2-2\epsilon}k_{\perp}}{(2\pi)^{3-2\epsilon}}\frac{k^2_{\perp}}{D^2_1
  D_2}.
\end{align}
To calculate these integrals, we   perform the Feynman
parametrization, then integrate over $k^0$ by using the residue
theorem. After that, we integrate over the transverse momentum
$k_{\perp}$ in $2-2\epsilon$ dimensions. At last, the Feynman
parameter $y$ is integrated out.

When integrating over $k_{\perp}$ the following formula is useful:
\begin{align}
  \int\frac{d^{D}k_{\perp}}{(2\pi)^D}\frac{1}{(k^2_{\perp}+\Delta)^n}=\frac{1}{(4\pi)^{\frac{D}{2}}}\frac{\Gamma\left(n-\frac{D}{2}\right)}{\Gamma(n)}\frac{1}{\Delta^{n-\frac{D}{2}}}.
\label{eq:dimreg}
\end{align}
We should note that this formula is valid only when
$\mathrm{Re}\left(n-D/2\right)>0$. For a linearly divergent
integral, $n-D/2=-1/2<0$, this formula is not valid.
However, under the spirit of analytical continuation, one can assign
a finite value to a linearly divergent integral by using
Eq.~\eqref{eq:dimreg}. With such analytical continuation, the
linearly divergent integrals $I_1$ and $I_4$ are finite in
the dimensional regularization scheme. Therefore, the final result of
Fig.~\ref{fig:gluonreal}(a) is finite as well. We believe that these results are only meaningful when a renormalization procedure is performed. This will be studied in a
forthcoming work.

By evaluating these integrals, we have
\begin{align}
 I_1&=-\frac{i P^z}{4\pi}\left\{ \begin{aligned}
&x^3\ln\frac{x-1}{x}+x^2+\frac{x}{8}+\frac{1}{8(x-1)}+\frac{11}{24},~~~~&x>1\\
&x^3 \ln\frac{(1-x+x^2)m^2_g}{4x(1-x)(P^z)^2}-\frac{(x-1) x^4 (P^{z})^2}{2 \left(1-x+x^2\right) m_g^2}\\
&+\frac{64 x^8-216 x^7+374 x^6-394 x^5+252 x^4-110 x^3+43 x^2-24 x+8}{24 (x-1) (1-x+x^2)^2},   &0<x<1\\
&x^3\ln\frac{x}{x-1}-x^2-\frac{x}{8}-\frac{1}{8(x-1)}-\frac{11}{24},~~~~&x<0
\end{aligned}\right.
\\
 I_2&=-\frac{i}{16\pi (P^z)^2 }\left\{ \begin{aligned}
&\ln\frac{x-1}{x}+\frac{2x-1}{2x(x-1)},~~~~&x>1\\
&\ln\frac{4x(1-x)(P^z)^2}{(1-x+x^2)m^2_g}-\frac{2(x-1)x (P^z)^2}{(1-x+x^2)m^2_g}+\frac{2x^6-6x^5+10x^4-10x^3+2x^2+2x-1}{2x(x-1)(1-x+x^2)^2},   &0<x<1\\
&-\ln\frac{x-1}{x}-\frac{2x-1}{2x(x-1)},~~~~&x<0
\end{aligned}\right.
\\
 I_3&=-\frac{i}{32\pi (P^z)^3 }\left\{ \begin{aligned}
&\frac{1}{x^2(x-1)},~~~~&x>1\\
&-\frac{4 (x-1)(P^z)^2}{m^2_g (1-x+x^2)}+\frac{2 x^4-6 x^3+4 x^2-2 x+1}{(x-1) x^2 \left(1-x+x^2\right)^2},   &0<x<1\\
&-\frac{1}{x^2(x-1)},~~~~&x<0
\end{aligned}\right.
\\
 I_4&=\frac{i P^z}{12\pi }\left\{ \begin{aligned}
&3x-1,~~~~&x>1\\
&2x^3-3x^2+3x-1,   &0<x<1\\
&-(3x-1),~~~~&x<0
\end{aligned}\right.
\\
 I_5&=-\frac{i }{8\pi P^z}\left\{ \begin{aligned}
&(x-1)\ln\frac{x-1}{x}+1,~~~~&x>1\\
&(x-1)\ln\frac{(1-x+x^2)m^2_g}{4x(1-x)(P^z)^2}+2x-1,   &0<x<1\\
&-(x-1)\ln\frac{x-1}{x}-1,~~~~&x<0
\end{aligned}\right.
\end{align}
With these scalar integrals and Eq.~\eqref{eq:app:dimreg}, finally
we arrive at Eq.~\eqref{eq:gluonreal1:result}. Similar approach can
be applied to other diagrams. The complete results have been
presented in Sec.~\ref{sec:1loop}.

\section{Regularization with offshellness}
\label{sec:appendix_offshell}

In this appendix, we present the results in which offshellness of
external particle is introduced to regularize the collinear
divergence. In this scheme, quark and gluon are taken to be
massless, but the external quark or gluon line is off-shell with a
small negative offshellness $P^2<0$. The collinear divergence in
quasi and light-cone distributions is then regularized as   terms
proportional to $\ln(-P^2)$. One can also find  that the matching
coefficients in offshellness scheme is the same to the results in
Sec.~\ref{sec:1loop}~E, which indicates that the factorization is
independent of IR regulator. These results might  be useful for the
renormalization of quasi PDFs on the Lattice, for instance  in the RI/MOM
scheme.

\subsection{Quark in quark}

For quark quasi and light-cone PDFs, we have
\begin{eqnarray}
\t f_{q/q}^{(1)}(x,
P^z)\bigg\vert_{\mathrm{Fig}.\ref{fig:quarkreal}(a)}=\frac{\alpha_s
C_F}{2\pi}\left\{\begin{aligned}
&(x-1)\ln\frac{x-1}{x}+1,~~&x>1\\
&(x-1)\ln\frac{-P^2}{4(P^z)^2},~~&0<x<1\\
&(x-1)\ln\frac{x}{x-1}-1,~~&x<0
\end{aligned}
\right.
\end{eqnarray}
and
\begin{eqnarray}
f_{q/q}^{(1)}(x,\mu)\bigg\vert_{\mathrm{Fig}.\ref{fig:quarkreal}(a)}=\frac{\alpha_s
C_F}{2\pi}\left\{
\begin{aligned}
&0,~~&x>1,~~x<0\\
&(x-1)\left(\ln\frac{-P^2 x(1-x)}{{\mu}^2}\right)+x-2,~~&0<x<1
\end{aligned}
\right.
\end{eqnarray}
\begin{eqnarray}
&&\t f_{q/q}^{(1)}(x,
P^z)\bigg\vert_{\mathrm{Fig}.\ref{fig:quarkreal}(b)}=\t
f_{q/q}^{(1)}(x,
P^z)\bigg\vert_{\mathrm{Fig}.\ref{fig:quarkreal}(c)} =\frac{\alpha_s
C_F}{4\pi}\left\{
\begin{aligned}
&\frac{2x}{1-x}\ln\frac{x}{x-1}-\frac{1}{1-x},~~~~&x>1\\
&\frac{2x}{1-x}\ln\frac{4(P^z)^2}{-P^2}+\frac{1-2x}{1-x},~~~~&0<x<1\\
&\frac{2x}{1-x}\ln\frac{x-1}{x}+\frac{1}{1-x},~~~~&x<0
\end{aligned}\right.
\end{eqnarray}
and
\begin{eqnarray}
&&f_{q/q}^{(1)}(x,\mu)\bigg\vert_{\mathrm{Fig}.\ref{fig:quarkreal}(b)}=f_{q/q}^{(1)}(x,\mu)\bigg\vert_{\mathrm{Fig}.\ref{fig:quarkreal}(c)}
=\frac{\alpha_s C_F}{2\pi}\left\{
\begin{aligned}
&0,~~~~&x>1,~x<0\\
&\frac{x}{1-x}\ln\frac{\mu^2}{-P^2 x(1-x)},~~~~&0<x<1
\end{aligned}
\right.
\end{eqnarray}

\begin{eqnarray}
\t f_{q/q}^{(1)}(x,
P^z)\bigg\vert_{\mathrm{Fig}.\ref{fig:quarkreal}(d)}=\frac{\alpha_s
C_F}{2\pi}\left\{
\begin{aligned}
&\frac{1}{1-x},~~~~&x>1\\
&\frac{1}{x-1},~~~~&0<x<1\\
&\frac{1}{x-1},~~~~&x<0
\end{aligned}
\right.
\end{eqnarray}
and
\begin{eqnarray}
f_{q/q}^{(1)}(x,\mu)\bigg\vert_{\mathrm{Fig}.\ref{fig:quarkreal}(d)}=0.
\end{eqnarray}

\begin{eqnarray}
&&\delta\widetilde{\mathcal{C}}^{(1)}_{qq}\bigg\vert_{\mathrm{Fig}.\ref{fig:quarkvirtual}(a)}
=-\frac{\alpha_s C_F}{2\pi}\int dy\left\{\begin{aligned}
&(y-1)\ln\frac{y-1}{y}+1,~~&y>1\\
&(y-1)\ln\frac{-P^2}{4(P^z)^2},~~&0<y<1\\
&(1-y)\ln\frac{y-1}{y}-1,~~~~&y<0
\end{aligned}\right.
\end{eqnarray}
and
\begin{eqnarray}
\delta
{\mathcal{C}}^{(1)}_{qq}\bigg\vert_{\mathrm{Fig}.\ref{fig:quarkvirtual}(a)}=-\frac{\alpha_s
C_F}{2\pi}\int^1_0 dy~\left[(y-1)\left( \ln\frac{-P^2
y(1-y)}{\mu^2}\right)+y-2\right].
\end{eqnarray}

\begin{eqnarray}
&&\delta\widetilde{\mathcal{C}}^{(1)}_{qq}\bigg\vert_{\mathrm{Fig}.\ref{fig:quarkvirtual}(b)}
=-\frac{\alpha_s C_F}{2\pi}\int dy\left\{
\begin{aligned}
&\frac{2y}{1-y}\ln\frac{y}{y-1}-\frac{1}{1-y},~~&y>1\\
&\frac{2y}{1-y}\ln\frac{4(P^z)^2}{-P^2}+\frac{1-2y}{1-y},~~&0<y<1\\
&\frac{2y}{1-y}\ln\frac{y-1}{y}+\frac{1}{1-y},~~&y<0
\end{aligned}\right.
\end{eqnarray}
and
\begin{eqnarray}
\delta
\mathcal{C}^{(1)}_{qq}\bigg\vert_{\mathrm{Fig}.\ref{fig:quarkvirtual}(b)}=-\frac{\alpha_s
C_F}{2\pi}\int^1_0 dy  \frac{y}{1-y}\ln\frac{\mu^2}{-P^2y(1-y)}.
\end{eqnarray}

\begin{eqnarray}
\delta\widetilde{\mathcal{C}}^{(1)}_{qq}\bigg\vert_{\mathrm{Fig}.\ref{fig:quarkvirtual}(c)}=-\frac{\alpha_s
C_F}{2\pi}\left\{
\begin{aligned}
&\frac{1}{1-y},~~~~&y>1\\
&\frac{1}{y-1},~~~~&0<y<1\\
&\frac{1}{y-1},~~~~&y<0
\end{aligned}
\right.
\end{eqnarray}
and
\begin{eqnarray}
\delta
\mathcal{C}^{(1)}_{qq}\bigg\vert_{\mathrm{Fig}.\ref{fig:quarkvirtual}(c)}=0.
\end{eqnarray}

\subsection{Gluon in gluon}

For gluon quasi and light-cone PDFs, we have
\begin{align}
\tilde f_{g/g}^{(1)}(x,
P^z)\bigg\vert_{\mathrm{Fig}.\ref{fig:gluonreal}(a)}&=\frac{\alpha_s
C_A}{2\pi x}\left\{ \begin{array} {ll}\left(2 x^3-3 x^2+2 x-2\right)
\ln \frac{x-1}{x}+2 x^2-\frac{5 x}{2}+\frac83\ , & x>1
\\ \left(2 x^3-3
x^2+2 x-2\right)\ln \frac{- P^2}{4(P^z)^2}+\frac23 x^2 (8x-9)+\frac{13x}{2}-\frac83\ , &0<x<1\\
-\left(2 x^3-3 x^2+2 x-2\right) \ln \frac{x-1}{x}-2 x^2+\frac{5
x}{2}-\frac83\ ,&x<0\end{array}
\right.\label{eq:gluonreal1:offshell}
\end{align}
and
\begin{align}
 f_{g/g}^{(1)}(x,
\mu)\bigg\vert_{\mathrm{Fig}.\ref{fig:gluonreal}(a)}&=\frac{\alpha_s
C_A}{2\pi x}\left\{ \begin{array} {ll}  0 & x>1\  \mbox{or}\ x<0
\\ \left(2 x^3-3 x^2+2
x-2\right) \ln \frac{-P^2 x(1-x)}{\mu^2}+2x^3-2x^2+3x-2\ ,&0<x<1
\end{array} \right.
\end{align}

\begin{eqnarray}
\tilde f_{g/g}^{(1)}(x,
P^z)\bigg\vert_{\mathrm{Fig}.\ref{fig:gluonreal}(b)}=\tilde
f_{g/g}^{(1)}(x,
P^z)\bigg\vert_{\mathrm{Fig}.\ref{fig:gluonreal}(c)}=\frac{\alpha_s
C_A}{4\pi}\left\{
\begin{aligned}
&\frac{1}{1-x}\left[x(1+x)\ln\frac{x}{x-1}-2x+1\right],&x>1\\
&\frac{1}{1-x}\left[x(1+x)\ln\frac{4(P^z)^2}{-P^2}-(2x^2-2x+1)\right],&0<x<1\\
&\frac{1}{1-x}\left[x(1+x)\ln\frac{x-1}{x}+2x-1\right],&x<0
\end{aligned}\right.
\end{eqnarray}
and
\begin{eqnarray}
 f_{g/g}^{(1)}(x,
\mu)\bigg\vert_{\mathrm{Fig}.\ref{fig:gluonreal}(b)}=\tilde
f_{g/g}^{(1)}(x,
\mu)\bigg\vert_{\mathrm{Fig}.\ref{fig:gluonreal}(c)}=\frac{\alpha_s
C_A}{4\pi}\left\{
\begin{aligned}
&0,~~~~&x>1,~x<0\\
&{\frac{x(1+x)}{1-x}\ln\frac{\mu^2}{-P^2 x(1-x)}},~~~~&0<x<1
\end{aligned}
\right.
\end{eqnarray}

\begin{eqnarray}
\tilde f_{g/g}^{(1)}(x,
P^z)\bigg\vert_{\mathrm{Fig}.\ref{fig:gluonreal}(d)}=\frac{\alpha_s
C_A}{2\pi}\left\{ \begin{aligned}
&\frac{x}{1-x},~~~~&x>1\\
&\frac{x}{x-1},~~~~&0<x<1\\
&\frac{x}{x-1},~~~~&x<0
\end{aligned}\right.
\end{eqnarray}
while the corresponding light-cone PDF
\begin{eqnarray}
 f_{g/g}^{(1)}(x,
\mu)\bigg\vert_{\mathrm{Fig}.\ref{fig:gluonreal}(d)}=0.
\end{eqnarray}

\begin{eqnarray}
\tilde f_{g/g}^{(1)}(x,
P^z)\bigg\vert_{\mathrm{Fig}.\ref{fig:gluonreal}(e)}=\frac{\alpha_s
C_A}{2\pi}\left\{
\begin{aligned}
&\frac12,~~~~&x>1\\
&\frac12,~~~~&0<x<1\\
&-\frac12,~~~~&x<0
\end{aligned}
\right.
\end{eqnarray}
and
\begin{eqnarray}
 f_{g/g}^{(1)}(x,
\mu)\bigg\vert_{\mathrm{Fig}.\ref{fig:gluonreal}(e)}=0.
\end{eqnarray}

For the virtual diagrams, we have
\begin{align}
 \delta \mathcal{C}^{(1)}_{gg}\bigg\vert_{\mathrm{Fig}.\ref{fig:gluonvirtual}(a)}=\frac{\alpha_s}{4\pi}\left[\left(\frac53 C_A-\frac43 T_F n_f\right)\ln\frac{\mu^2}{-P^2}+\frac{31}{9}C_A-\frac{20}{9}T_F
  n_f\right]
\end{align}
which is the same to the result in Ref.~\cite{Ji:2005nu}. For quasi
distribution, we have
\begin{align}
  \delta\widetilde{\mathcal{C}}^{(1)}_{gg}\bigg\vert_{\mathrm{Fig}.\ref{fig:gluonvirtual}(a)}=\frac{\alpha_s}{4\pi}\left[\left(\frac53 C_A-\frac43 T_F n_f\right)\ln\frac{(P^z)^2}{-P^2}+\frac{31}{9}C_A-\frac{20}{9}T_F
  n_f\right].
\end{align}

\begin{eqnarray}
&&\delta\t
{\mathcal{C}}^{(1)}_{gg}\bigg\vert_{\mathrm{Fig}.\ref{fig:gluonvirtual}(b)}
=-\frac{\alpha_s C_A}{2\pi}\int dy \left\{
\begin{aligned}
&\frac{1+y}{1-y}\ln\frac{y}{y-1}+\frac{1-2y}{1-y},~~&y>1\\
&\frac{1+y}{1-y}\ln\frac{4(P^z)^2}{-P^2}-\frac{2y^2-2y+1}{1-y},~~&0<y<1\\
&\frac{1+y}{1-y}\ln\frac{y-1}{y}+\frac{2y-1}{1-y},~~&y<0
\end{aligned}\right.
\end{eqnarray}
for quasi distribution. For standard PDF we have
\begin{eqnarray}
\delta
{\mathcal{C}}^{(1)}_{gg}\bigg\vert_{\mathrm{Fig}.\ref{fig:gluonvirtual}(b)}={-\frac{\alpha_s
C_A}{2\pi}\int^1_0 dy \frac{1+y}{1-y}\ln\frac{\mu^2}{-P^2 y(1-y)}}.
\end{eqnarray}

\begin{eqnarray}
\delta\t
{\mathcal{C}}^{(1)}_{gg}\bigg\vert_{\mathrm{Fig}.\ref{fig:gluonvirtual}(c)}=-\frac{\alpha_s
C_A}{2\pi}\int dy\left\{\begin{aligned}
&\frac{y}{1-y},~~~~&y>1\\
&\frac{y}{y-1},~~~~&0<y<1\\
&\frac{y}{y-1},~~~~&y<0
\end{aligned}\right.
\end{eqnarray}
The standard light-cone PDF has
\begin{eqnarray}
\delta
\mathcal{C}^{(1)}_{gg}\bigg\vert_{\mathrm{Fig}.\ref{fig:gluonvirtual}(c)}=0.
\end{eqnarray}

\subsection{Gluon in quark and quark in gluon}
For quark-to-gluon splitting functions, we have

\begin{eqnarray}
\widetilde f_{g/q}(x, P^z)=\frac{\alpha_s C_F}{2\pi x}\left\{
\begin{aligned}
&(1+(1-x)^2)\ln\frac{x}{x-1}-x+\frac52,~~~~&x>1\\
&(1+(1-x)^2)\ln\frac{4 (P^z)^2}{-P^2}-x^2+4x-\frac52,~~~~&0<x<1\\
&(1+(1-x)^2)\ln\frac{x-1}{x}+x-\frac52,~~~~&x<0
\end{aligned}\right.
\end{eqnarray}
\begin{eqnarray}
f_{g/q}(x,\mu)=\frac{\alpha_s C_F}{2\pi x}\left\{
\begin{aligned}
&0,~~~~&x>1,~~x<0\\
&\left(1+(1-x)^2\right)\ln\frac{\mu^2}{-P^2x(1-x)}-x^2+x-2,~~&0<x<1
\end{aligned}
\right.
\end{eqnarray}
and for gluon-to-quark splitting functions, we have
\begin{eqnarray}
\widetilde f_{q/g}(x,P^z)=\frac{\alpha_s T_F}{2\pi}\left\{
\begin{aligned}
&(x^2+(1-x)^2)\ln\frac{x}{x-1}-2x+1,~~~~&x>1\\
&(x^2+(1-x)^2)\ln\frac{4(P^z)^2}{-P^2}-4x^2+2x,~~~~&0<x<1\\
&(x^2+(1-x)^2)\ln\frac{x-1}{x}+2x-1.~~~~&x<0
\end{aligned}\right.
\end{eqnarray}

\begin{eqnarray}
f_{q/g}(x,\mu)=\frac{\alpha_s T_F}{2\pi}\left\{
\begin{aligned}
&0,~~~~&x>1,~~x<0\\
&(x^2+(1-x)^2)\ln\frac{\mu^2}{-P^2 x(1-x)}-1.~~~~&0<x<1
\end{aligned}
\right.
\end{eqnarray}


\end{document}